\documentclass[notitlepage,twocolumn,prl,nobalancelastpage,amssymb,superscriptaddress,showpacs,longbibliography]{revtex4-1}
\usepackage{graphicx}
\usepackage{amsmath}
\usepackage{siunitx}
\usepackage{amssymb}
\usepackage[colorlinks=true,citecolor=blue,linkcolor=blue]{hyperref}
\usepackage{float}
\usepackage{comment}
\usepackage[dvipsnames]{xcolor}
\usepackage{physics}
\usepackage{commath}
\usepackage{soul}

\newcommand{\bk}{\boldsymbol{k}}
\newcommand{\bs}[1]{\boldsymbol{#1}}
\newcommand{\be}{\boldsymbol{e}}
\newcommand{\bE}{\boldsymbol{E}}

\newcommand{\bj}{\boldsymbol{j}}

\newcommand{\brho}{\boldsymbol{\rho}}

\begin{document}

\title{Catastrophes, Optical Multistabilities, and Chiral Photocurrent Hysteresis in Driven Weyl Semimetals}

\author{Christopher Yang}
\address{Department of Physics, IQIM, California Institute of Technology, Pasadena, CA 91125, USA}
\address{Department of Physics and Astronomy, University of California, Irvine, Irvine, California 92697, USA}

\author{Gil Refael}
\address{Department of Physics, IQIM, California Institute of Technology, Pasadena, CA 91125, USA}

\author{Frederik Nathan}
\affiliation{NNF Quantum Computing Programme and Center for Quantum Devices, 
Niels Bohr Institute, University of Copenhagen, Copenhagen 2100, Denmark}

\begin{abstract}

The unique band topology of Weyl semimetals provides them with a diverse array of strong photovoltaic response phenomena. Here we explore how plasma screening affects these phenomena. We find that the screening field is highly nonlinear, can contain topological contributions, and can amplify the photoresponse, with the self-consistent response exhibiting the hallmarks of bistability, including hysteresis and catastrophes. The bistability emerges from a crossover from linear to saturated polarization as a function of the field amplitude,  because carriers in a Weyl cone have bounded group speed $v_F$.  %
When time-reversal symmetry is broken, tilted Weyl nodes add a Berry-curvature contribution to the in-plane current, making the nonlinear screening response helicity selective. %
\end{abstract}

\maketitle

\textit{Introduction}---The linear dispersion and unique band topology of Weyl semimetals (WSMs) produce exotic electronic responses with  promising applications in optics, magnetism, and electronics~\cite{Nagaosa2020,RevModPhys.90.015001,PhysRevLett.107.127205,Xu2015,PhysRevX.5.031013,Lu2015,Zhang2017,PhysRevLett.119.196403,Vafek2014,PhysRevLett.128.176401,PhysRevB.95.041104,PhysRevB.89.245121,PhysRevB.99.075137,PhysRevB.98.195446,Nathan_2022,deJuan2017,Ma2017,Ji2019,PhysRevB.97.195151,Osterhoudt_2019,Guo2023,PhysRevResearch.6.013048,PhysRevB.94.245121,PhysRevB.97.241118,Ji2019,PhysRevB.98.195446,Ji2020,Lai2022,Wu2016,Almutairi2020,Cheng2020,PhysRevB.101.165426,PhysRevB.102.165417,PhysRevB.96.085114,Yang2022,Moore2018,Li2023,PhysRevB.105.144435,Swekis2021,Destraz2020,PhysRevLett.128.176401,Zhang2022,PhysRevLett.131.186704,Araki2019,PhysRevB.109.L121108,Nielsen1983,PhysRevB.88.104412,PhysRevLett.109.181602,Liu2019,Xiong2015,Zhang2016,Ong2021,PhysRevB.92.041203,PhysRevX.6.041021,PhysRevB.87.245131,PhysRevB.102.235134,PhysRevLett.124.176402,Mikitik2019,Burkov2015,Wang2017,Gorbar2018,Hu2019,PhysRevB.100.201102,PhysRevLett.111.027201,PhysRevB.105.174406,PhysRevB.96.205443}. 
While some effects are associated with interband photoabsorption~\cite{deJuan2017,Ma2017,Ji2019,PhysRevB.97.195151,Osterhoudt_2019,Guo2023,PhysRevResearch.6.013048,PhysRevB.94.245121,PhysRevB.97.241118}, WSMs also support unique responses  from {\it intraband} dynamics~\cite{PhysRevLett.115.216806,PhysRevB.93.201202,PhysRevB.99.075150,PhysRevB.94.235123,Nathan_2022,Mikhailov_2008,PhysRevB.101.085424,PhysRevResearch.2.043252} when the driving is adiabatic with respect to the excitation gap. These include topological charge- or energy pump effects~\cite{PhysRevLett.117.216601,PhysRevB.95.245211,PhysRevB.103.L201105,Nathan_2022}. %

A key question is how screening modifies these exotic effects. WSMs have plasma frequencies in the THz regime and generically screen low-frequency radiation, but their unique dispersion may lead to unconventional screening effects.

\begin{figure}[t!]
    \centering
    \includegraphics[width=0.98\linewidth]{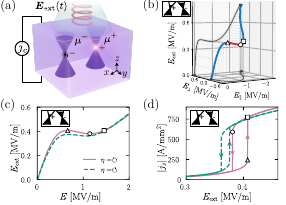}
    \caption{\textbf{{Multistable optical response and photocurrent} in a driven WSM.}
    (a) Circularly polarized light drives WNs of differing chemical
    potential and tilt in a WSM with broken time-reversal and inversion
    symmetry; the in-plane current polarizes the sample, and the induced
    field modifies the internal field amplitude $E$.
    (b) Self-consistent steady states of a minimal two-node model under a
    right-circularly polarized field; $E_\parallel$, $E_\perp$ are the
    {in-plane} components of the internal field parallel and perpendicular to
    $\boldsymbol E_{\rm ext}(t)$, and blue (red) branches are stable
    (unstable).
    (c) External versus internal amplitude
     for left (dashed) and right (solid) circular
    polarization. Over a window of weak $E_{\rm ext}$, three
    solutions coexist and the internal field is
    amplified. Triangle and circle mark the two folds; increasing $E_{\mathrm{ext}}$ from zero, $E$ jumps from triangle to square.
    (d) {Hysteretic} out-of-plane photocurrent. Increasing and decreasing $E_{\mathrm{ext}}$ 
    trace different loops for the two helicities.}
    \label{fig:intro}
\end{figure}

Motivated by this question, we analyze screening in WSMs driven adiabatically by circularly polarized light at frequencies comparable to the plasma frequency. We find that the strongly nonlinear in-plane response from the WN can produce an optical multistability: a weak external drive becomes amplified inside the material [Figs.~\ref{fig:intro}(b)--(c)], and the response exhibits the hallmarks of multistabilities, including hysteresis and catastrophes. %
The threshold external amplitude for the multistability can be much smaller in WSMs than for other nonlinear dispersions, since the nonlinear photoresponse is visible at all field strengths for sufficiently small chemical potentials. This effect is generic for all classes of WSMs.

When inversion symmetry is broken, the bistability affects the charge pumping and dc photocurrent effects from Refs.~\cite{PhysRevLett.117.216601,PhysRevB.95.245211,PhysRevB.103.L201105}. Here we show that {the strong enhancement of the internal field leads to} %
charge pumping at external fields weaker than anticipated from the unscreened response,
and that  the photocurrent inherits the catastrophe and hysteresis, see Fig.~\ref{fig:intro}(d).

We finally report that WSMs with tilted WNs and broken time-reversal symmetry can exhibit
a helicity-selective plasmonic response, %
in which the anomalous velocity amplifies one helicity of the optical field while leaving the opposite helicity suppressed.

\textit{Single-node model.}---Let us first introduce our model. %
Focusing on the conduction and valence bands, the Hamiltonian near a WN reads $H(\boldsymbol{k}) = \xi \hbar v_F \boldsymbol{k} \cdot \boldsymbol{\sigma} + \hbar \boldsymbol{V} \cdot \boldsymbol{k}$ \cite{RevModPhys.90.015001}.
Here, $\boldsymbol{\sigma}$ is the vector of Pauli matrices in the orbital basis, and $\boldsymbol{k}$ is measured from the WN at $\boldsymbol{k} = 0$. The Fermi velocity $v_{F}$ sets the gap opening rate away from the WN, $\boldsymbol{V}$ parameterizes a tilt, and $\xi \in \{ +1,-1\}$ is the \textit{chirality} of the WN.
We let $\varepsilon_{\boldsymbol{k}\nu}$ and $|\psi_{\boldsymbol{k}\nu} \rangle$ denote the energies and eigenstates of ${H}(\boldsymbol{k})$,
with $\nu =0$ ($\nu=1$)
the valence (conduction) band.
The WN is a point source of the Berry curvature ${\bs \Omega_\nu(\bk)}\equiv   \nabla_{\bk} \times \langle \psi_{\boldsymbol{k}\nu} | i \nabla_{\bk}  | \psi_{\boldsymbol{k}\nu}\rangle $, with $\nabla \cdot \bs{\Omega}_\nu (\bk) = 2\pi \xi (-1)^\nu \delta(\boldsymbol{k})$, and $\delta(\boldsymbol{k})$ the Dirac delta function. %

We consider coherent circularly polarized light traveling along the $z$ axis and driving the system at frequency $f$, with polarization $\eta = \circlearrowleft$ or $\circlearrowright$ denoting left- or right-circular polarization (LCP or RCP). The internal vector potential is $\boldsymbol{A}(t) = A(\cos \omega t, c_{\eta}\sin \omega t, 0)$, with angular frequency $\omega = 2\pi f$, $c_{\circlearrowleft} = 1$, $c_{\circlearrowright} = -1$, and internal electric field $\boldsymbol{E}(t) = -\partial_t \boldsymbol{A}(t)$. Electrons couple via $\boldsymbol{k} \to \boldsymbol{k}  + e \boldsymbol{A}(t)/\hbar$, with $e$ the magnitude of the electron charge, giving the time-periodic Hamiltonian ${H}(\boldsymbol{k},t) = {H}(\boldsymbol{k} + e \boldsymbol{A}(t)/\hbar)$. We consider the {\it adiabatic regime} $hf \ll v_FeA$ \cite{Adiabatic}.

The strong anomalous velocity of the Weyl semimetal produces significant nonlinear responses. Here we focus on a dispersion-insensitive out-of-plane dc photocurrent realizing Thouless' adiabatic charge pump~\cite{PhysRevB.103.L201105}.
To see its origin, consider the electron modes at fixed $xy$-plane momentum $\boldsymbol{k}_{\perp} =(k_x,k_y)$ in a charge-neutral WN. These states form a one-dimensional fermionic chain along $z$. If $|\boldsymbol{k}_{\perp}|<eA/\hbar$, the bands of the Bloch Hamiltonian ${H}(\boldsymbol{k},t)$
for this chain remain fully gapped over a full driving cycle, forming a closed loop around the WN.
This adiabatic trajectory realizes a Thouless charge pump \cite{thouless-pump}, transporting one electron per cycle in the $z$-direction: $\mathcal I^z(\boldsymbol{k}_{\perp}) =- c_{\eta} \mathcal C_{\boldsymbol{k}_{\perp}}ef$. Here, $\mathcal C _{\boldsymbol{k}_{\perp}}$ is the net charge of Weyl nodes inside the closed cylinder $\mathcal S$ defined by $\bk'(t)\equiv \bk+e{\bs A}(t)/\hbar$ over one driving period $0\leq t\leq T_{\mathrm{dr}}$, where $T_{\mathrm{dr}}=1/f$, at fixed $\boldsymbol{k}_{\perp}$. %
Summing over electronic momentum-space trajectories enclosing the WN gives the total pumped current: $-c_{\eta} ef$ per chain times the number of chains, i.e., the reciprocal-space disk area $\pi(eA/\hbar)^2$ divided by the chain area $(2\pi)^2/\mathcal A$, where $\mathcal A$ is the $x$-$y$ cross section of the WSM. Dividing by $\mathcal A$ gives the topological current density 
\begin{equation}
    j^z _{\rm top}=- c_{\eta} \xi \frac{ef}{4\pi}  \left(\frac{eA}{\hbar}\right)^2,
\end{equation}
as first presented in Refs.~\cite{PhysRevLett.117.216601,PhysRevB.95.245211,PhysRevB.103.L201105}.
We analyze multiple nodes, imperfect filling, and anisotropy below.

\textit{Plasma screening---}{At moderate intensities, this photocurrent can reach $j^z _{\rm top}\sim {\rm A}/{\rm mm}^2$ for $E\sim 10^5\, {\rm V}/{\rm m}$ and $f\sim {\rm THz}$.} We investigate whether such amplitudes %
are realistic under plasma screening. %
\begin{figure}
    \includegraphics[width=\columnwidth]{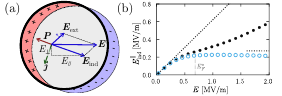}
    
    \caption{\textbf{Origin of the screening effect.} (a) In-plane cross section of the oblate disk, showing the screening response and relevant vector quantities.
   (b) Induced field parallel to the internal electric field, $E_{\rm ind}^{\parallel}$, vs. internal amplitude $E$ for an untilted WN. Black: full master equation result {[Eq.~\eqref{eq:dm_lindblad}, End Matter]}. Blue: finite-doping contribution $E_{\rm ind}^{\parallel} - E_{\rm ind}^{\parallel}|_{\mu=0}$, obtained by subtracting the nonadiabatic interband response of a charge-neutral WN. Black dashed lines: semiclassical asymptotes; vertical line: $E_F^*=\mu\omega/(ev_F)$.}
    \label{fig:polarization_sketch}
\end{figure}

We model the WSM as an oblate spheroidal disk
with its symmetry axis along the propagation direction.   {Light-induced plasma oscillations create the surface-charge oscillation in Fig.~\ref{fig:polarization_sketch}(a), inducing a field ${\boldsymbol E}_{\rm ind}(t)$ that modifies the internal field $\boldsymbol{E}(t)$ relative to the external field outside the WSM, $\boldsymbol{E}_{\text{ext}}(t)$, via $\boldsymbol{E}(t)={\boldsymbol E}_{\rm ind}(t)+\boldsymbol{E}_{\text{ext}}(t)$. }
For the spheroidal geometry considered here, $\boldsymbol E_{\rm ind}$ is approximately uniform when the sample is homogeneous. This field is determined by the light-induced current $\boldsymbol j(t)$ that results in a uniform polarization $\boldsymbol P(t)$, with
\begin{equation}\label{eq:screenrel}
    \partial_t\boldsymbol P(t)
    =\boldsymbol j(t)
    ,
    \qquad
    \boldsymbol E_{\rm ind}(t)
    =-\frac{N_\parallel\boldsymbol P(t)}{\epsilon_0}.
\end{equation}
Here $N_\parallel=\frac{1}{2}[1-\frac{1+\zeta^2}{\zeta^3}(\zeta-\tan^{-1}\zeta)]$, with $\zeta=\sqrt{(D/d)^2-1}$, where $D$ and $d$ are the in-plane diameter and maximum thickness, respectively~\cite{PhysRev.67.351}.  We neglect currents arising from the intra-unit-cell motion of electrons, such as the Bloch-state magnetization~\cite{PhysRevLett.116.077201}, and retain only the transport current associated with intercell motion across the lattice.
In the adiabatic regime, we can compute the transport current $\boldsymbol j(t)$ using a semiclassical
approximation~\cite{Adiabatic,Nathan_2022},
\begin{align} \label{eq:current}
    \boldsymbol j(t) &= -e \int \frac{d^3 \boldsymbol k}{(2\pi)^3} \sum_\nu g_\nu(\boldsymbol k,t)\boldsymbol v_\nu(\boldsymbol k,t), \\ 
    \boldsymbol{v}_{\nu}(\boldsymbol{k},t) &\equiv \frac{ \nabla_{\boldsymbol{k}}}{\hbar} \varepsilon_{\boldsymbol{k}\nu}(t) - e \boldsymbol{\Omega}_{\nu}\left(\boldsymbol{k} + \frac{e}{\hbar} \boldsymbol{A}(t)\right) \times \dfrac{\boldsymbol{E}(t)}{ \hbar}.
\label{eq:phasevel}
\end{align}
The terms in Eq. (\ref{eq:phasevel}) are the group velocity and Berry-curvature-induced anomalous velocity, while $g_\nu(\boldsymbol{k},t)$ %
 denotes the occupation of band $\nu$. %

{For simplicity, we assume the band model is rotationally symmetric about $z$. (The End Matter shows that moderately breaking this symmetry %
leaves our conclusions unchanged.) We therefore take $\boldsymbol j(t)$ to respect the combined time-translation and rotational symmetry and have the form %
\begin{equation}
    \bj(t)
    =\bar j_z \hat z
    +j_P(E)\hat\be(t)
    +j_N(E)\hat z\times\hat\be(t),
    \label{eq:j_decomp}
\end{equation}
where $\hat\be(t)=\bE(t)/E$. Here $\bar j_z$ is the adiabatic photocurrent along the propagation direction, %
while the in-plane components $j_P(E)$ and $j_N(E)$ generate the screening response.}

{We calculate $j_{P,N}$ using complementary semiclassical and numerical master equation approaches. The semiclassical treatment provides an analytic description in the slow-relaxation regime $\tau f\gg1$, where $\tau$ is the electronic scattering time. Here the occupation is approximately given by the time average of the instantaneous Fermi sea, $g_{\nu}(\boldsymbol{k},t)\approx \frac{1}{T_{\mathrm{dr}}}\int_t^{t+T_{\mathrm{dr}}}du,f_{\nu}^{\mathrm{eq}}(\boldsymbol{k},u)$, where $f_{\nu}^{\mathrm{eq}}$ is the instantaneous equilibrium occupation of band $\nu$~\cite{Nathan_2022}. 

The complementary Lindblad master equation analysis models electrons via their momentum-resolved reduced density matrix ${\rho}(\boldsymbol{k},t)$, incorporating a phenomenological relaxation toward the instantaneous thermal state at rate $1/\tau$ (see End Matter). Throughout our simulations, we use realistic experimental parameters $\tau=0.9\ \mathrm{ps}$ and $f=1.09\ \mathrm{THz}$ ($\hbar\omega=4.5\ \mathrm{meV}$)~\cite{PhysRevB.95.085202,PhysRevB.95.041104}. Our results below show that the semiclassical analysis agrees with the master equation down to relatively fast ($\tau \simeq 1/f$) relaxation times. By retaining the full time dependence and interband coherences, the Lindblad master equation also captures nonadiabatic corrections absent from Eq.~\eqref{eq:phasevel}. 

We compute the screening response for a realistic sample geometry modeled as an oblate spheroid with diameter $D=21.5 \ \mu\mathrm{m}$ and thickness $d=0.15 \ \mu\mathrm{m}$, corresponding to $N_\parallel=5.43\times10^{-3}$. These dimensions are much smaller than the $275 \ \mu\mathrm{m}$ wavelength of the applied field. The sample thickness is also below the THz skin depth of $\sim 1 \ \mu\mathrm{m}$~\cite{Matus2022}, yet sufficiently large for the system to remain quasi-3D~\footnote{For a WN with $\mu=50 \ \mathrm{meV}$, $\mu d/(\pi\hbar v_F)=7.3$; this value is even larger when the tilt is oriented along the $z$ direction.}}.

\begin{figure}[t!]
    \centering
    \includegraphics[width=0.98\linewidth]{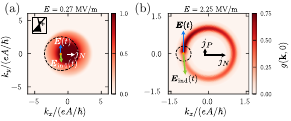}
    \caption{\textbf{Origin of the nonlinear in-plane response.}
    Steady-state occupation $g(\boldsymbol{k},0)$ for $\tau f \gtrsim 1$ in the $\nu = 1$ conduction band at $k_z=0$. Orange dot: instantaneous WN position. Black dashed circle: instantaneous equilibrium Fermi contour. Vectors: oscillating current components $j_{P}$, $j_{N}$ and induced field $\boldsymbol{E}_{\text{ind}}(t)$. (a) Weak field, $eA/\hbar\ll k_F$: the drive weakly smears and displaces the Fermi sea. %
    (b) Strong field, $eA/\hbar \gg k_F$: the distribution smears into a ring of radius $eA/\hbar$ centered off the WN.
    }
    \label{fig:ferms}
\end{figure}

\begin{figure}[t!]
    \centering
    \includegraphics[width=0.98\linewidth]{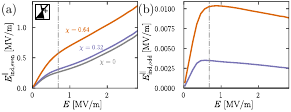}
    \caption{\textbf{Screening from a single WN.}
    Helicity-even and helicity-odd components of the induced field, $E^{\parallel}_{\rm ind,even/odd} = \frac{1}{2}(E_{\mathrm{ind}}^{\parallel}|_{\eta = \circlearrowleft} \pm E_{\mathrm{ind}}^{\parallel}|_{\eta = \circlearrowright})$, for tilts $\chi=0$ (gray), $0.32$ (purple), $0.64$ (orange); vertical lines mark $E_F^*$. (a) The even response, from the helicity-independent group-velocity current, is linear at weak fields and strongly sublinear beyond $E_F^*$. (b) The Berry-curvature-induced odd component is finite only for tilted WNs ($\chi \neq 0$).}
    \label{fig:inplane}
\end{figure}

\textit{Screening from an untilted isolated WN.}---We first consider an isolated, isotropic Weyl node. {The simulations in Figs.~\ref{fig:ferms} and~\ref{fig:inplane} use $v_F = 5 \times 10^{5} \ \mathrm{m/s}$, $\mu = 50 \ \mathrm{meV}$, $\xi = +1$, and $T = 20 \ \mathrm{K}$.} Because the in-plane anomalous velocity in Eq.~\eqref{eq:phasevel} is odd in $k_z$, %
the screening current comes solely from the group-velocity term.  For $eA/\hbar\ll k_F$, the drive weakly displaces the Fermi sphere by $eA/\hbar=eE/(\hbar\omega)$ [see Fig.~\ref{fig:ferms}(a)], producing the ordinary Drude response.  This gives $j_N=-D_W c_\eta E/\omega$, where the Weyl Drude weight $D_W=e^2\nu(\mu)\langle v_x^2\rangle_{\rm FS}=e^2v_Fk_F^2/(6\pi^2\hbar)$ follows from the density of states $\nu(\mu)=k_F^2/(2\pi^2\hbar v_F)$ and the average $\langle v_x^2\rangle_{\rm FS}=v_F^2/3$ across the Fermi surface.  Thus, {we obtain a conventional linear response}
\begin{equation} 
    j_P=0,\quad
    j_N\simeq -c_\eta\frac{e v_F k_F^2}{6\pi^2} \frac{eA}{\hbar} ,
    \quad \mathrm{when} \ \frac{eA}{\hbar}\ll k_F.
    \label{eq:inplane_weak}
\end{equation}

In the strong-amplitude limit $eA/\hbar\gg k_F$, the time-averaged occupation is instead a narrow torus [see Fig.~\ref{fig:ferms}(b)]. The Fermi volume is smeared uniformly around the circular WN trajectory of radius $eA/\hbar$, centered at the equilibrium position $\bk = 0$.  Most occupied states lie along $c_\eta \hat{\boldsymbol z}\times \boldsymbol{\hat{e}}(t)$ from the WN
and therefore have velocity projection along this direction.
More quantitatively, a point at angular position $\varphi$ around the torus has group velocity along the chord connecting it to the WN, at angle $\varphi/2$ to $+\hat x$.  Angular averaging gives $\langle v_x\rangle=(2\pi)^{-1}\int_{-\pi}^{\pi}d\varphi\,v_F\cos(\varphi/2)=2v_F/\pi$, while the projection parallel to $\hat\be$ averages to zero.  Thus $j_N=-c_\eta e n\langle v_x\rangle$, with $n=k_F^3/(6\pi^2)$ denoting the occupied density, and
\begin{equation}
    j_P=0,\quad
    j_N\simeq -c_\eta\frac{e v_F k_F^3}{3\pi^3},
    \quad \text{when} \ \frac{eA}{\hbar}\gg k_F .
    \label{eq:inplane_strong}
\end{equation}
{Thus, in the strong-amplitude regime, this analysis suggests the current {\it saturates}:} Increasing the field amplitude %
does not increase the average group velocity of carriers, which is fixed by $v_F$. %

The polarization of the system is the key quantity that gives rise to the nonlinear screening effect. In the slow-relaxation limit, $j_P = 0$, so the current is perpendicular to the internal electric field. Thus, the polarization and induced field are aligned to the internal field [Eq.~\eqref{eq:screenrel}], so $\boldsymbol{E}_{\mathrm{ind}}={E}^{\parallel}_{\mathrm{ind}}\hat{\boldsymbol e}(t)$.  Using Eqs.~\eqref{eq:inplane_weak} and \eqref{eq:inplane_strong},
$
    {E}^{\parallel}_{\mathrm{ind}}\simeq \frac{N_\parallel e^2v_Fk_F^2}{6 \epsilon_0 \pi^2\hbar\omega^2}E$ for $ E\ll E_F^*$, and $
    {E}^{\parallel}_{\mathrm{ind}}\simeq \frac{N_\parallel ev_Fk_F^3}{3 \epsilon_0 \pi^3\omega}$ for $ E\gg E_F^*$, where $
 E_F^*\equiv\frac{\mu\omega}{ev_F}$.
The first is ordinary linear plasma screening. The second is the nonlinear plateau characteristic of Weyl dispersions. Fig.~\ref{fig:polarization_sketch}(b) compares these estimates with {numerical simulations}. %
The semiclassical analysis agrees closely with the blue markers showing the contribution to the induced field from finite doping $E_{\rm ind}^{\parallel} - E_{\rm ind}^{\parallel}|_{\mu=0}$, which subtracts the nonadiabatic interband response of a charge-neutral WN.
The total induced field (black points) retains this sharp nonlinearity for $E > E_F^*$, with clear sublinear scaling and a slope change.

{\textit{Screening from a tilted isolated WN.}}---{We now consider screening with a finite tilt along the $z$-axis, %
$\bs V=V_z\hat z$ with $V_z = \chi v_F$ (other orientations are discussed in the End Matter).
For a type-I cone, the tilted Fermi pocket %
is an ellipsoid centered at $-\chi k_{\rm F}/(1-\chi^2) \boldsymbol{\hat{z}}$ with eccentricity $|\chi|$.
The primary effect of the tilt at order $\chi$} is a net in-plane anomalous velocity. Because this velocity is proportional to $-\Omega_{\nu,z}\hat z\times\bE(t)$, the displaced Fermi volume prevents cancellation of $\Omega_z\propto k_z/|\bk|^3$ between $\pm k_z$. The resulting helicity-dependent current corrects $E_{\mathrm{ind}}^{\parallel}$ by $-\xi c_\eta
    \frac{N_\parallel e^2\chi k_F}{12\pi^2\epsilon_0\hbar\omega}E$ for $ E\ll E_F^*$ and $-\xi c_\eta\frac{N_\parallel e\chi k_F^2}{8\pi^3\epsilon_0}$ for $
     E\gg E_F^*$
to leading order in $\chi$ [see the SM~\cite{SupplementalMaterials} and Ref.~\cite{PhysRevLett.119.036601} for the weak-field result]. The tilt also modifies the group-velocity terms of $E_{\mathrm{ind}}^{\parallel}$: at weak field it renormalizes the in-plane Drude weight by $F_\chi=
\frac{3[
2\chi-(1-\chi^2)
\ln\!(\frac{1+\chi}{1-\chi})
]}
{4\chi^3(1-\chi^2)}$ (see the SM), whereas the strong-field current scales with the Fermi volume, which increases by a factor $(1-\chi^2)^{-2}$.

Fig.~\ref{fig:inplane} shows the tilt dependence of the helicity-even  and odd components of $E_{\mathrm{ind}}^{\parallel}$, {obtained from  our numerical simulations} %
for a single node. The correction above explains the helicity-odd contribution in panel (b). The tilt-induced enhancement of the helicity-even response in panel (a) is consistent with the enlarged Fermi volume. Crucially, the onset of the strong-field nonlinearity is essentially unchanged even for large tilt.

{\textit{Multi-node effects and symmetry considerations}.---We now sum the screening contributions from all WNs to obtain the total component of the induced field parallel to the internal field $\bE(t)$,}
\begin{equation}
    {E}^{\parallel}_{\mathrm{ind}}\simeq \left\{\begin{array}{ll}
     \sum_i\left[\frac{N_\parallel e^2v_Fk_{F,i}^2 F_{\chi_i}}{6\epsilon_0 \pi^2\hbar\omega^2}%
    -\frac{ \xi_i c_\eta N_\parallel e^2\chi_i k_{F,i}}{12\pi^2\epsilon_0\hbar\omega}\right] E, & E\ll E_l^* \\
   \sum_i\left[\frac{N_\parallel ev_{F}k_{F,i}^3}{3\pi^3\epsilon_0\omega(1-\chi_i^2)^2}%
    -\frac{\xi_i c_\eta N_\parallel e\chi_i k_{F,i}^2}{8\pi^3\epsilon_0}\right], & E\gg E_u^* 
    \end{array}\right.
    \label{eq:two_node_Eind}
\end{equation}
where $i$ indexes the WN with tilt $\chi_i$, chemical potential $\mu_i = \hbar v_F k_{F,i}$, and chirality $\xi_i$, and $E_l^* = \mathrm{min}[\{\mu_i \omega/(e v_F)\}]$ and $E_u^* = \mathrm{max}[\{\mu_i \omega/(e v_F)\}]$. This relates the induced field to the internal field magnitude $E$. 

{The Hamiltonians for each WN can be related to each other by inversion symmetry (IS) or time-reversal symmetry (TRS). For WSMs with TRS, the tilt-independent group velocity contribution in Eq.~\eqref{eq:two_node_Eind} adds constructively, while the helicity- and chirality-odd induced field cancels. Thus, ${E}^{\parallel}_{\mathrm{ind}}$ is helicity independent yet retains a nonlinear saturation at large $E$. In a WSM with broken TRS, the helicity-odd induced field can be nonzero, producing helicity-dependent screening.}

{\textit{Optical multistability.}---}We now close the feedback loop between the driven electrons and the electromagnetic field, relating the external field $E_{\rm ext}$ to the internal amplitude $E$ controlling the quantities above.  Writing $\bE_{\rm ind}(t)=E_{\rm ind}^{\parallel}(E)\hat\be(t)+E_{\rm ind}^{\perp}(E)\hat z\times\hat\be(t)$, self-consistency $\bE(t)=\bE_{\rm ext}(t)+\bE_{\rm ind}(t)$ gives
\begin{equation}
    |E_{\rm ext}|^2=
    \left[E-E_{\rm ind}^{\parallel}(E)\right]^2
    +\left[E_{\rm ind}^{\perp}(E)\right]^2 .
    \label{eq:steady_roots}
\end{equation}
In the slow-relaxation regime studied here, $E_{\rm ind}^{\perp}$ is small, so $E_{\rm ind}^{\parallel}$ controls the response.  At low field,
$E_{\rm ind}^{\parallel}\propto E$.  %
For a proportionality constant exceeding $1$, the response overscreens the applied field.
At larger $E$, the sublinearity of the polarization [reduced $d E^{\parallel}_{\mathrm{ind}}/dE$, see Fig.~\ref{fig:polarization_sketch}(b)] cuts off this growth.
Eq.~\eqref{eq:steady_roots} then admits three internal amplitudes per $E_{\rm ext}$ within a finite interval. {Because the large-field sublinearity of $E^{\parallel}_{\mathrm{ind}}$ occurs with or without the helicity-dependent component in Eq.~\eqref{eq:two_node_Eind}, the optical multistability should arise in all classes of WSMs.}

{To display the full array of effects, including helicity-dependent screening, we consider in Figs.~\ref{fig:intro}(b) and (c) the lowest-symmetry case: a WSM with broken IS and TRS. We use a minimal model with two WNs of opposite chiralities, $\mu_1=50 \ \mathrm{meV}$, $\mu_2=45 \ \mathrm{meV}$, and $\chi_{1,2}=\pm0.64$.} The helicity-dependence in Fig.~\ref{fig:intro}(c) reflects the anomalous velocity contribution to the induced field.
We assess the stability of each solution by linearizing the field dynamics around small perturbations to the internal vector potential. (See End Matter for details.) On the central branch (red in Fig.~\ref{fig:intro}(b)), a {perturbation $\delta A$ evolves as %
\(\delta A(t)\propto e^{\lambda t}\), with \(\operatorname{Re}\lambda>0\), so the fixed point is unstable.} All perturbations are instead damped on the two outer branches (blue).

\begin{figure}[t!]
    \centering
    \includegraphics[width=0.98\linewidth]{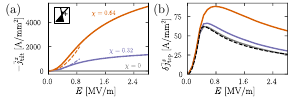}
    \caption{\textbf{Photocurrent from a single WN.}
    (a) Helicity-independent and (b) helicity-dependent dc photocurrents along $z$, for tilts $\chi=0$ (gray), $0.32$ (purple), $0.64$ (orange). Panel (b) plots $\delta\bar j_{\rm top}^z\equiv\bar j_{\rm top}^z-\bar j_{\rm top}^z|_{\mu = 0}$, isolating the finite-$\mu$ contribution since $\bar j_{\rm top}^z|_{\mu = 0}$ cancels between opposite-chirality WNs in a realistic system. Dashed colored segments in (a) are the weak-field estimate of Eq.~\eqref{eq:jztilt}; the black dashed curve in (b) is the $\chi=0$ semiclassical result of Eq.~\eqref{eq:jtop}.}
    \label{fig:photocurrents}
\end{figure}

\textit{Effects on the photocurrent---}
The photocurrent along the light propagation inherits the multistability and hysteresis of the internal field amplitude.
Away from charge neutrality, the anomalous velocity must be weighted by the occupation function. For $\tau f \gg 1$, this gives $\bar{j}_{\eta}^z  = \sum_i [\bar  j_{{\rm top},i}^z + \bar j_{{\rm tilt},i}^z]$, where \begin{align}
\bar j^z_{{\rm top},i} &= -c_\eta \xi _i \frac{ef}{4\pi} \left(\frac{eA}{\hbar}\right)^2 \mathcal F(\mu_i / (e v_F A))
\label{eq:jtop}\\ 
    \bar j^z_{{\rm tilt},i}
    &\simeq
    -\chi_i\frac{e\mu_i}{15\pi^2\hbar}\left(\frac{eA}{\hbar} \right)^2,
    \qquad eA/\hbar\ll k_{F,i},
\label{eq:jztilt}
\end{align}
and $\mathcal F(x)$ is an $\mathcal O(1)$ function decreasing from unity as $1- \frac{x^3}{3\pi}[\ln(4/x)+\frac{1}{3} ]+O(x^5)$ for $\mu \ll v_FeA$, while saturating at $1/3$ for $\mu \geq  2v_FeA$ (see full expression in the SM).
The second term reflects the preferential population of states with momentum along the tilt: under the adiabatic drive, their instantaneous energy stays below the chemical potential for a longer fraction of the period, leaving a net current along the tilt. {After summing over all WNs, a finite topological current requires broken IS, while the tilt current requires broken TRS. We compute both numerically for the minimal two-node configuration described below Eq.~\eqref{eq:steady_roots} and used in Fig.~\ref{fig:intro}.} Fig.~\ref{fig:photocurrents} extracts $\bar j^z_{\rm tilt}=(\bar j^z_{\circlearrowleft}+\bar j^z_{\circlearrowright})/2$ and $\bar j^z_{\rm top}=(\bar j^z_{\circlearrowleft}-\bar j^z_{\circlearrowright})/2$ for comparison with the semiclassical limits above. The steep field dependence produces strong photocurrent signatures of the multistability [Fig.~\ref{fig:intro}(d)].

\textit{Conclusions}---
We have shown that the in-plane intraband current of a circularly driven WSM produces nonlinear local-field screening. The polarization crosses over from the ordinary linear plasma regime, $P\propto E$, to sublinear scaling once $eA/\hbar$ exceeds the Fermi momentum. {This crossover occurs generically in WSMs with or without inversion or time-reversal symmetry.} {Remarkably, in magnetic WSMs with broken time-reversal symmetry and net anisotropy along laser propagation, the Berry-curvature renders the screening helicity selective.}   Treated self-consistently, these nonlinearities generate field amplification, bistability, and hysteresis at low threshold amplitudes set by the chemical potential. The enhanced internal field produces large photocurrents $\sim 100 \ \mathrm{A/mm^2}$. 

The strong-field intraband multistability studied here differs from weak-field effects in thin-film WSMs~\cite{vxms-qwmf}, bistability driven by self-induced Berry flux~\cite{Rudner2019}, and bistability in cavities or photonic crystals containing 3D Dirac materials or WSMs~\cite{Long_22,HE2024416171}. Related studies of optical bistability in graphene employed a weak-field expansion~\cite{PhysRevB.90.125425} and phenomenological models extrapolated to large field amplitudes~\cite{PhysRevB.92.121407,Cox2014}, with subsequent work underscoring the need for a nonperturbative analysis~\cite{PhysRevB.95.085432}. Our nonperturbative treatment connects strong-field bistability to the bounded group velocity of the Dirac Hamiltonian, extending the effect to a broad class of WSMs that can also host a new interplay between Berry curvature and nonlinear screening.

The multistability effects are observable for a WSM in a thin-disk geometry with thickness $\sim 0.1 \ \mathrm{\mu m}$ below the skin depth, Fermi velocity $v_F \sim 5 \times 10^{5} \ \mathrm{m/s}$, external fields $\sim 0.1$--$1 \ \mathrm{MV/m}$, and frequencies $\sim 1 \ \mathrm{THz}$ near the bare plasma resonance. {Together, the screening effects enable amplification and control of adiabatic Weyl photocurrents in the THz regime.}

\begin{acknowledgments}
{\it Acknowledgements---}We thank Ivar Martin, Cyprian Lewandowski, Elio K\"{o}nig, Tobias H\"{o}lder, Takahiro Morimoto, Alexander Tyner, Luis Jauregui, Thomas Scaffidi, Javier Sanchez-Yamagishi, Filippo Capolino, and Hiro Ishizuka for insightful discussions. C.Y. gratefully acknowledges support from the DOE NNSA Stewardship Science Graduate Fellowship program, which is provided under cooperative agreement number DE-NA0003960, the Eddleman Quantum Institute postdoctoral fellowship, and the Moore Foundation postdoctoral fellowship. G.R. is grateful for support from the Simons Foundation and the Institute of Quantum Information and Matter, as well as support from the NSF DMR grant number 1839271. This work is supported by ARO MURI Grant No. W911NF-16-1-0361, and was performed in part at Aspen Center for Physics, which is supported by National Science Foundation grant PHY-1607611. F.N. was supported by the U.S. Department of Energy, Office of Science, Basic Energy Sciences under award DE-SC0019166, the Simons Foundation under award 623768, and the Carlsberg Foundation, grant CF22-0727.
\end{acknowledgments}

\bibliography{references_use.bib}

\onecolumngrid

\newpage
\begin{center}
\ \vskip 0.2cm
{\large\bf End Matter}
\end{center}
\twocolumngrid

\textit{Lindblad master equation}.---Under the drive, the density matrix ${\rho}(\bk,t)$ is affected by incoherent processes such as electron-electron and electron-phonon collisions, which we model phenomenologically through the master equation
\begin{equation}
 \partial_t\rho(\bk,t)
 =\frac{i}{\hbar}[\rho(\bk, t), H(\bk,t)]
 -\frac{1}{\tau}\left[\rho(\bk,t)-\rho^{\rm eq}(\bk,t)\right],
 \label{eq:dm_lindblad}
\end{equation}
Here, we assume that electrons relax to a thermal state in the instantaneous eigenbasis of ${H}(\bk,t)$. This is physically justified when the correlation time of the environment is fast relative to the driving frequency \cite{PhysRevB.102.115109}. The equilibrium density matrix is given by $\rho^{\rm eq}(\bk,t)=\sum_{\nu=0,1}f_{\rm FD}[\epsilon_{\bk \nu}(t)-\mu]P_{\bk \nu}(t)$, where $f_{\rm FD}(\varepsilon)=[e^{\varepsilon /(k_BT)}+1]^{-1}$ and $P_{\bk \nu}(t)$ projects onto band $\nu$ of $H(\bk,t)$.

To solve the master equation numerically, we expand $\rho(\bk,t)$ {in terms of its Fourier coefficients, $\rho(\bk,t)=\sum_{|n|\leq N_{\rm F}}\rho_{n}({\bk })e^{in\omega t}$, with $N_{\rm F}$  a frequency cutoff.
We expand $\rho^{\rm eq}(t)$ and $H(\bk,t)$ similarly, using the same cutoff. 
}
{We collect the Fourier coefficients in a vector as $\brho({\bk})=(\rho_{-N_{\rm F}}(\bk),\ldots \rho_{\rm NF}(\bk))$ and define $\brho_{\rm eq}(\bk)$ likewise. 
The master equation then becomes a time-independent linear equation,} %
$\mathcal M({\bk})\boldsymbol\rho({\bk})=\boldsymbol \brho_{\rm eq}({\bk})$, with $\mathcal M_{nm}({\bk})=(1+in\omega\tau)\delta_{nm}+(i\tau/\hbar){\rm{ad}}_{H_{n-m}(\bk)}$ and  ${\rm ad}_{\mathcal O}[\rho]=[\mathcal O,\rho]$. The linear equation can be solved straightforwardly, e.g., using the matrix representations of the superoperators involved.
We then compute the current from 
\begin{equation} \label{eq:currenteq}
    \boldsymbol{j}_{\eta}(t) = -e \int \frac{d^3 \boldsymbol{k}}{(2\pi)^3} \mathrm{Tr}[{\rho}(\bk,t)\hbar^{-1} \nabla_{\bk} {H}(\bk,t)],
\end{equation}
{approximating the integral using discrete sampling of $\bk$-points, and }leveraging the azimuthal symmetry \footnote{Writing the momentum as $\boldsymbol{k} = (k_{\perp} \cos\phi, k_{\perp} \sin\phi, k_z)$, we note that the Hamiltonian has the symmetry $H(k_{\perp},\phi, k_z,t) = U(\phi) H(k_{\perp},0,k_z,t-c_\eta\phi/\omega) U^\dagger(\phi)$, with $U(\phi)=e^{-i\phi\sigma_z/2}$, which is inherited by the instantaneous thermal density matrix ${\rho}^{\mathrm{eq}}$ and by extension all the time-periodic steady states ${\rho}$ of the master equation under the relaxation-time approximation. Thus the numerical algorithm only solves for the density matrix on the plane $k_y = 0$ and $k_x \geq 0$, and uses $U(\phi)$ to infer the result at other momenta related by an azimuthal rotation.} for computational efficiency \footnote{In the numerical calculations, the momentum integral is evaluated over a finite square domain. If this domain is not centered on the instantaneous position of the Weyl node, the finite integration bound produces a large spurious current. We therefore evaluate the in-plane current via  Eq.~\eqref{eq:currenteq} by replacing the the density matrix according to ${\rho}(\bk,t) \to {\rho}(\bk,t)-{\rho}^{\mathrm{eq}}(\bk,t)$, since the equilibrium distribution generates no net current. The dc $z$-current in Fig.~\ref{fig:photocurrents}(b) is instead reported as the difference relative to charge neutrality: $\delta\bar j^z_{\rm top}=\bar j^z_{\rm top}-\bar j^z_{\rm top}|_{\mu =0}$. The subtracted $\bar j^z_{\rm top}|_{\mu =0}$ contribution cancels between opposite-chirality nodes in a realistic WSM.}.

\textit{Stability analysis.}---To assess the stability of a given steady state, we hold the external drive fixed, perturb the internal vector potential by $\delta\boldsymbol A(t)$, and determine its subsequent dynamics. This induces corresponding perturbations $\delta\boldsymbol P(t)$ and $\delta\boldsymbol\rho(\bk,t)$ of the polarization and density matrix, governed by the Maxwell relations [Eq.\eqref{eq:screenrel}] and the Lindblad master equation [Eq.~\eqref{eq:dm_lindblad}]. Linearizing around the steady state gives a closed system of linear   equations for  $\delta\boldsymbol X(t)=(\{\delta\rho(\bk,t)\},\delta\boldsymbol P(t),\delta\boldsymbol A(t))$ of the form $\partial_t\delta\boldsymbol X(t)=\mathcal L(t)\delta\boldsymbol X(t)$, where $\mathcal L(t+T_{\mathrm{dr}})=\mathcal L(t)$; see the Supplemental Material (SM) for a full expression for $\mathcal L(t)$. By Floquet's theorem, its eigenmodes can be written as $\delta\boldsymbol X(t)=e^{\lambda t}\boldsymbol u(t)$, where $\boldsymbol u(t+T_{\mathrm{dr}})=\boldsymbol u(t)$. The Hermiticity of the Hamiltonian and the Maxwell equations endow {$\mathcal L(t)$} with the symmetry {$\mathcal{C} \mathcal{L}(t) = \mathcal{L}(t) \mathcal{C}$}, where $\mathcal{C}$ denotes complex conjugation of $\delta \boldsymbol{A}$ and $\delta \boldsymbol{P}$ and Hermitian conjugation of $\delta \rho(\bk,t)$. Thus $\lambda^*$ is also an eigenvalue. The physical perturbation is given by $\delta \boldsymbol{X}_{\mathrm{phys}}(t) = e^{\lambda t} \boldsymbol{u}(t) + e^{\lambda^* t} \mathcal{C} \boldsymbol{u}(t)$, with a decay or growth rate given by $\operatorname{Re}\lambda$.

\begin{figure}[!t]
    \centering
    \includegraphics[width=0.95\columnwidth]{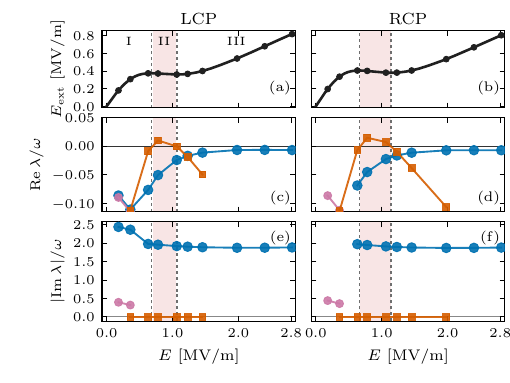}
    \caption{\textbf{Self-consistent response and its stability.} Left/right columns: LCP/RCP drive. (a),(b) External field $E_{\rm ext}$ versus internal field amplitude $E$, with black dots marking the sampled amplitudes used for the linear stability analysis. (c),(d) Real and (e),(f) nonnegative imaginary parts of selected roots in the box $-0.115\leq\operatorname{Re}\lambda/\omega\leq0.05$: the high-frequency oscillatory plasma-mode (blue circles), the real amplitude mode (orange squares), and a lower-frequency damped pair (pink circles). Lines connect only successive sampled roots shown in the box. The amplitude mode crosses into $\operatorname{Re}\lambda>0$ in region II, rendering it unstable, while both oscillatory pairs remain damped.}
    \label{fig:density_matrix_stability}
\end{figure}

In practice, determining $\lambda$ requires solving a Floquet eigenvalue problem for {$\mathcal L(t)$, a large matrix} spanning the density-matrix degrees of freedom at every momentum point, making direct numerical diagonalization impractical. Instead, we {can effectively integrate out} $\delta\rho(\bk,t)$ by computing the current response to linear order in $\delta\boldsymbol A(t)$. Working in the Sambe space, we write the internal field perturbation as $\delta\boldsymbol A(t)=e^{\lambda t}\sum_{n=-K}^{K}
      \boldsymbol a_n e^{-in\omega t}$.
      {Note that the multiple harmonics  allow our analysis to account for high-harmonic generation effects (see next subsection).} We can obtain the current response as $\delta\boldsymbol j_{\perp,n}
    =\sum_{m=-K}^{K}\boldsymbol G_{nm}(\lambda)\boldsymbol a_m .$
Here $\boldsymbol G_{nm}(\lambda)$ is the {Fourier-decomposed} susceptibility obtained from the linearized Lindblad equation, which {encodes the current response to a vector potential perturbation, including retardation effects. Crucially, we can  compute this quantity numerically [see the SM for details]~}\footnote{At $\lambda=in\omega$, the $n$th harmonic of the vector potential is
spatially uniform and time independent. In the continuum, this corresponds to a gauge shift and can produce no physical current. Numerical integration on a finite-sized grid, however, can weakly violate this identity. We restore it by
replacing
$[\boldsymbol G(\lambda)]_{m n}\to
[\boldsymbol G(\lambda)]_{m n}
-[\boldsymbol G(in\omega)]_{m n}$.}. Writing $\delta\boldsymbol P(t)=e^{\lambda t}\sum_n\boldsymbol p_n e^{-in\omega t}$, %
the Maxwell relations $\partial_t\delta\boldsymbol A=\frac{N_{\parallel}}{\epsilon_0}\delta\boldsymbol P$ 
and $\partial_t\delta\boldsymbol P=\delta\boldsymbol j_\perp$ give $(\lambda-in\omega)\boldsymbol a_n =\frac{N_{\parallel}}{\epsilon_0}\boldsymbol p_n$ and $(\lambda-in\omega)\boldsymbol p_n=\sum_mG_{nm}(\lambda)\boldsymbol a_m$, which combine to give the nonlinear eigenvalue problem
\begin{equation}
    (\lambda-in\omega)^2 \boldsymbol a_n=\frac{N_{\parallel}}{\epsilon_0} \sum_m G_{nm}(\lambda)\boldsymbol a_m .
\label{eq:charpoly}
\end{equation}
We solve this equation %
using Newton--Raphson iteration and use winding-number counts based on the argument principle to verify that all roots within the chosen contours have been accounted for. A root with $\operatorname{Re}\lambda>0$ represents an instability, whereas $\operatorname{Re}\lambda<0$ represents decay.

When the system has continuous azimuthal symmetry, with no anisotropy in the plane of polarization, the system responds in only the fundamental harmonic $n=\pm1$. The problem can therefore be reduced to $K=1$.

Fig.~\ref{fig:density_matrix_stability} tracks roots within the box $-0.115\leq\operatorname{Re}\lambda/\omega\leq0.05$  as a function of the internal field amplitude. Regions I, II, and III correspond respectively to the lower positive-slope branch, the backward-bending negative-slope branch, and the upper positive-slope branch of the multistability curve [panel (a)]. We focus on the high-frequency oscillatory complex-conjugate pair (blue circles), associated with the plasma mode of the disk, and the nonoscillatory amplitude mode (orange squares). The amplitude mode crosses into $\operatorname{Re}\lambda>0$ in Region II [panels (c) and (d)], producing an instability. Both oscillatory pairs remain damped at the sampled points. The high-frequency plasma-mode pair becomes less damped with increasing internal field amplitude, while its oscillation frequency [panels (e) and (f)] decreases due to the suppression of the Drude weight at strong fields. Roots outside the sampled box are not shown.

\begin{figure}[t!]
    \centering
    \includegraphics[width=\columnwidth]{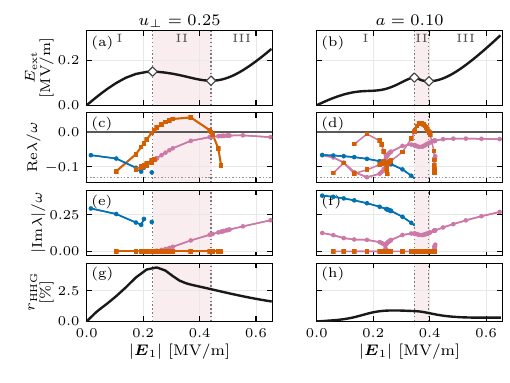}
    \caption{\textbf{Robustness of the multistability to broken azimuthal symmetry.} Semiclassical results for two routes of broken axial symmetry in the polarization plane: transverse tilt ($u_\perp=0.25$, $a=0$; left) and velocity anisotropy ($a=0.10$, $u_\perp=0$; right), with contacts only along $z$ and hence no dc in-plane current. Rows, top to bottom: self-consistent response $E_{\rm ext}(|\boldsymbol{E}_1|)$, with $\boldsymbol{E}_1$ the fundamental harmonic inside the WSM; $\operatorname{Re}\lambda/\omega$ and $|\operatorname{Im}\lambda|/\omega$ of the least-damped modes from the multitone stability analysis; and the harmonic content $r_{\rm HHG}\equiv {[\sum_{m=2}^{L}|\mathbf{E}_{m}|^2}/{|\mathbf{E}_{1}|^2}]^{1/2}$. Orange: amplitude modes driving the region-II instability; blue and pink: least-damped oscillatory modes. We exclude modes with decay rates faster than the phenomenological relaxation rate $1/\tau$.}
    \label{fig:anisotropy_hhg}
\end{figure}

\textit{Effect of in-plane anisotropy.}---
The main text assumed rotational symmetry about the axis of laser propagation. We now test the robustness of the multistability when this symmetry is broken in the plane of polarization, using a semiclassical analysis with the Weyl Hamiltonian $H(\mathbf{k})
=\hbar\mathbf{w}\cdot\mathbf{k}\,I
+\xi\hbar\boldsymbol{\sigma}\cdot \mathsf V \mathbf{k}$, where $\xi=\pm1$, $\mathsf V$ is the velocity matrix, and $\mathbf w$ the tilt velocity. We take $\mathsf V =\operatorname{diag}(v_x,v_y,v_z)$, characterize the in-plane velocity anisotropy by $a=(v_x-v_y)/(v_x+v_y)$, and define the dimensionless tilt $\mathbf u=\mathsf V^{-T}\mathbf w$ ($u_i=w_i/v_i$), parameterized as $\mathbf u=(u_\perp\cos\phi,u_\perp\sin\phi,u_z)$ with $u_z$ the tilt along the propagation axis and $u_\perp$ the in-plane tilt.

Breaking the axial symmetry lets the internal field become elliptically polarized and develop higher harmonics. In an isolated sample with contacts only along $z$, it also generates an in-plane dc field $\mathbf{E}_{\mathrm{dc}}$ (but no dc current). We therefore write $\mathbf E(t)=\mathbf{E}_{\mathrm{dc}} +\sum_{m=1}^{K}
[\mathbf E_m e^{-im\omega t}+ \mathrm{c.c.}]$. For each field, we evaluate the current semiclassically as described below and extract its Fourier components, $\mathbf J_\ell[\{\mathbf E_m\}]
=\frac{1}{T_{\mathrm{dr}}}\int_0^{T_{\mathrm{dr}}}dt\,
\mathbf j(t;\{\mathbf E_m\})e^{i\ell\omega t}$ with $T_{\mathrm{dr}}=\frac{2\pi}{\omega}$. The external field contains only the fundamental, $\mathbf E_{{\rm ext},1}=E_{\rm ext}(\hat{\mathbf x}+ic_{\eta}\hat{\mathbf y})/2$, with $\mathbf E_{{\rm ext},\ell>1}=0$ and $c_{\eta}=\pm1$ the helicity. The steady state solution follows from the self-consistency equations
\begin{equation}
\mathbf E_{{\rm ext},\ell} = \mathbf E_\ell+\frac{iN_\parallel}{\epsilon_0\ell\omega}
\mathbf J_\ell[\{\mathbf E_m\}].
\label{eq:anisotropic_maxwell}
\end{equation}
where $1 \leq \ell \leq K$. Together with $\mathbf{J}_0 =0$, this is a coupled set of $4K+2$ real nonlinear equations for the real and imaginary $x$ and $y$ components of $\mathbf E_m$.

To obtain $\mathbf J_\ell[\{\mathbf E_m\}]$ from the semiclassical equations of motion, Eqs.~\eqref{eq:current} and \eqref{eq:phasevel}, we insert the multiharmonic vector potential with $\mathbf A_\ell=\mathbf E_\ell/(i\ell\omega)$ for the ac harmonics and compute the occupation $g_n(\mathbf k,t)
=\int_{t-T_{\mathrm{dr}}}^{t}ds\,W_\tau(t-s)f_n^{\rm eq}(\mathbf k,s),$ using $W_\tau(\Delta t)
=\frac{e^{-\Delta t/\tau}}{\tau(1-e^{-T_{\mathrm{dr}}/\tau})}$ under the relaxation-time approximation. After numerical integration over $\bk$ and $s$, the time-dependent result is Fourier transformed to obtain $\mathbf J_\ell$ \footnote{We change variables to $\mathbf p=\mathsf V[\mathbf k+e\mathbf A(s)/\hbar]$. This transforms the integration domain into the time-independent tilted Fermi volume $|\mathbf p|+\mathbf u\cdot\mathbf p<\mu/\hbar$. The resulting three-dimensional integral over $\mathbf p$ and $s$ is evaluated numerically.}.

To separate the effects of velocity anisotropy from transverse tilt, we vary $a$ at $u_\perp=0$ and then $u_\perp$ at $a=0$, and quantify high harmonic generation by $r_{\rm HHG}\equiv
{[\sum_{m=2}^{L}|\mathbf{E}_{m}|^2}/{|\mathbf{E}_{1}|^2}]^{1/2}$. We determine the stability of the multitone states by solving the characteristic equation Eq.~(\ref{eq:charpoly}), keeping $K = L$ harmonics. 

Fig.~\ref{fig:anisotropy_hhg} compares a transverse tilt $u_\perp=0.25$ at $a=0$ (calculated using $K = 7$) with a velocity anisotropy $a=0.10$ at $u_\perp=0$ (calculated using $K = 11$); both exhibit an optical multistability. We use $\tau=1\,{\rm ps}$, $\hbar\omega=5\,{\rm meV}$, $N_\parallel=7.25\times10^{-3}$, $v_{\rm ref}=1.5\times10^6\,{\rm m/s}$, and two nodes with $(\mu,\xi,u_z)=(100\,{\rm meV},+1,0.6)$ and $(15\,{\rm meV},-1,-0.6)$. The HHG amplitude under the transverse tilt reaches $r_{\rm HHG}\approx4.36\%$, almost entirely second harmonic, while it reaches $0.88\%$ the velocity anisotropy, almost entirely third harmonic.

\end{document}


\widetext
\onecolumngrid

\begin{center}
\textbf{\large Supplemental Material:
\\Catastrophes, Optical Multistabilities, and Chiral Photocurrent Hysteresis in Driven Weyl Semimetals}
\\[0.4ex] Christopher Yang, Gil Refael, and Frederik Nathan
\end{center}
\par
\setcounter{page}{1}

\setcounter{equation}{0}
\setcounter{figure}{0}
\setcounter{table}{0}
\setcounter{page}{1}
\makeatletter
\renewcommand{\theequation}{S\arabic{equation}}
\renewcommand{\thefigure}{S\arabic{figure}}
\renewcommand{\thetable}{S\arabic{table}}
\setcounter{secnumdepth}{4}
\setcounter{tocdepth}{3}

{\hypersetup{linkcolor=black}
\tableofcontents
}
\vspace{1em}

\def\lf{\left\lfloor}   

\def\rf{\right\rfloor}

Here we provide the semiclassical calculation of the photocurrent $\boldsymbol j(t)$ quoted in the main text. {Before we proceed to the technical details of the derivation, which are detailed in the following sections, we begin by outlining the structure of the calculation.}

Our calculation of $\boldsymbol j(t)$ is based on solving the Lindblad master equation in Eq.~(11) of the End Matter,
\begin{equation} \label{eq:rellinb}
    \partial_t {{\rho}}(\boldsymbol{k},t) = \frac{i}{\hbar} [{\rho}(\boldsymbol{k},t), {H}(\boldsymbol{k},t)] - \frac{1}{\tau}[{{\rho}(\boldsymbol{k},t) - {\rho}^{\text{eq}}(\boldsymbol{k},t)}],
\end{equation}
with $ H(\bk,t)$, $ \rho(\bk,t)$, and $ \rho_{\rm eq}(\bk ,t)$ denoting the Bloch Hamiltonian, density matrix and equilibrium density matrix, respectively, while $\tau$ is the phenomenological relaxation time of the system (see main text for more details).
The photocurrent is obtained from the {steady-state solution to Eq. (\ref{eq:rellinb}), $ \rho_{\rm s}(\bk,t)$} via $\boldsymbol j(t)=-\frac{e}{(2\pi)^3}\int d^3\bk \text{Tr}[{\rho}_{\rm s}(\boldsymbol{k},t) \hbar^{-1} \nabla_{\boldsymbol{k}} {H}(\boldsymbol{k},t)]$ [see Eq.~(12) in the End Matter]. In the quasi-adiabatic limit that we consider, this integral can be approximated by~\cite{PhysRevResearch.4.043060}
\begin{equation}
    \boldsymbol j(t)\approx -e\sum_{\nu}\int\frac{d^3\boldsymbol{k}}{(2\pi)^3}g_{\nu}(\boldsymbol{k},t) \boldsymbol{v}_{\nu}(\boldsymbol{k},t).
    \label{eqa:approx_current}
\end{equation} 
Here  $g_\nu(\mathbf k,t)=\operatorname{Tr}\!\left[\rho_s(\mathbf k,t)P_{\nu\mathbf k}(t)\right]$ denotes  the steady-state occupation of the instantaneous electronic bands, with $P_{\nu\mathbf k}(t)$ denoting the projector onto the instantaneous eigenstate in band $\nu$ of $H(\bk,t)$.
Moreover,
\begin{equation}
    \boldsymbol{v}_{\nu}(\boldsymbol{k},t) \equiv \frac{ \nabla_{\boldsymbol{k}}}{\hbar} \varepsilon_{\boldsymbol{k}\nu}(t) - e \boldsymbol{\Omega}_{\nu}\left(\boldsymbol{k} + \frac{e}{\hbar} \boldsymbol{A}(t)\right) \times \dfrac{\boldsymbol{E}(t)}{ \hbar}\label{eqa:phasevel}
\end{equation}
denotes the phase velocity of the electrons in band $\nu$. Here, $\boldsymbol\Omega_\nu(\bk + e\boldsymbol{A}(t)/\hbar)$ and $\varepsilon_{\bk \nu}(t)$ are respectively the Berry curvature and eigenenergies of the bands of the Bloch Hamiltonian $H(\bk,t)$, $\boldsymbol{A}(t)$ is the drive-induced magnetic vector potential, and $\boldsymbol{E}(t) = -\partial_t \boldsymbol{A}(t)$  the corresponding electric field. {We provide more details on the quantities above in the main text.}

{The remainder of this supplement is devoted to calculating the photocurrent $\boldsymbol j(t)$ using the results above, and to analyzing the stability of the resulting self-consistent states.}
The photocurrent is controlled by two independent dimensionless parameters: the dimensionless tilt $\chi$ of the Weyl cone along the laser propagation axis, and the ratio $x_F\equiv k_F/(eA/\hbar)$ between the Fermi momentum and the drive-induced momentum displacement. We organize the calculation accordingly. Section~\ref{sec:occupation} sets up the model and derives the steady-state occupation function. Section~\ref{sec:untilted} then treats an isolated, untilted node ($\chi=0$), and Sec.~\ref{sec:tilt} the consequences of a finite longitudinal tilt ($\chi \neq 0$). In both Secs.~\ref{sec:untilted} and \ref{sec:tilt}, we first compute the dc photocurrent {along} the axis of laser propagation, and then the oscillating photocurrent {in the plane} of the laser field responsible for screening. Unlike the out-of-plane current, the in-plane current does not admit a closed form valid for all $x_F$, so it is treated separately in the strong-field (nonlinear) regime $k_F\ll eA/\hbar$, which is the regime responsible for the multistability, and in the weak-field regime $k_F\gg eA/\hbar$, which coincides with linear-response theory. Table~\ref{tab:roadmap} summarizes the results of each subsection and their relation to the results presented in the main text.

The final two sections address the stability of the steady state. In Sec.~\ref{sec:stability}, we perform linear-stability analysis using the Lindblad master equation, specializing to the case of Weyl nodes with azimuthal symmetry about the laser propagation axis.

\begin{table}[h]
    \centering
    \renewcommand{\arraystretch}{1.35}
    \begin{tabular}{l|l|l}
    \hline\hline
    & Untilted node, $\chi=0$ (Sec.~\ref{sec:untilted}) & Tilted node, $\chi\neq0$ (Sec.~\ref{sec:tilt})\\
    \hline
    \begin{tabular}[c]{@{}l@{}}Out-of-plane dc photocurrent $\bar j^z$\end{tabular}
    & \begin{tabular}[c]{@{}l@{}}Sec.~\ref{sec:zcurrsecuntilt}: exact for all $x_F$\\ $\to$ Eq.~(9) of the main text\end{tabular}
    & \begin{tabular}[c]{@{}l@{}}Sec.~\ref{sec:zcurrenttilt}: weak field, $O(\chi)$\\ $\to$ Eq.~(10) of the main text\end{tabular}\\
    \hline
    \begin{tabular}[c]{@{}l@{}}In-plane screening current $j_{P,N}$,\\ strong field $k_F\ll eA/\hbar$\end{tabular}
    & \begin{tabular}[c]{@{}l@{}}Sec.~\ref{sec:plasmon}\\ $\to$ Eq.~(6) of the main text\end{tabular}
    & \begin{tabular}[c]{@{}l@{}}Sec.~\ref{sec:inplanecurrtilt}\\ $\to$  Eq.~(7) of the main text \end{tabular}\\
    \hline
    \begin{tabular}[c]{@{}l@{}}In-plane screening current $j_{P,N}$,\\ weak field $k_F\gg eA/\hbar$\end{tabular}
    & \begin{tabular}[c]{@{}l@{}}Sec.~\ref{sec:weak_untilted}\\ $\to$ Eq.~(5) of the main text\end{tabular}
    & \begin{tabular}[c]{@{}l@{}}Sec.~\ref{sec:weak_tilt}\\ $\to$ Eq.~(7) of the main text \end{tabular}\\
    \hline\hline
    \end{tabular}
    \caption{Organization of Secs.~\ref{sec:occupation}--\ref{sec:tilt}. The two key parameters are the dimensionless tilt $\chi$ along the laser propagation axis and ratio of the Fermi momentum $k_F$ to the drive-induced momentum displacement $eA/\hbar$, denoted $x_F=k_F/(eA/\hbar)$.}
    \label{tab:roadmap}
\end{table}

\section{Model, notation, and steady-state electronic distribution} \label{sec:occupation}

\subsection{Model and notation}\label{sec:model}
Throughout Secs.~\ref{sec:occupation}--\ref{sec:tilt} we consider an isolated Weyl node with a tilt along the propagation ($z$) axis of the drive, described by the Hamiltonian
\begin{equation} \label{eq:tiltedh}
    H(\boldsymbol{k}) = {\xi} \hbar v_F \boldsymbol{k} \cdot \boldsymbol{\sigma}+\hbar V_z k_z\, I ,
\end{equation}
where $v_F$ is the Fermi velocity, $\xi=\pm1$ is the chirality of the node, $V_z$ is the tilt velocity, and $I$ is the $2\times2$ identity matrix. We use the dimensionless tilt parameter $\chi\equiv V_z/v_F$, and recover the untilted node for $\chi=0$. The node is driven by circularly polarized light propagating along $\hat z$, with in-plane vector potential $\boldsymbol A(t)$ of magnitude $A$ and electric field $\boldsymbol E(t)=-\partial_t\boldsymbol A(t)$ of magnitude $E=\omega A$, where $\omega=2\pi f$ and $f$ is the drive frequency. For circular polarization, we have $\boldsymbol E(t)=-c_\eta\omega\,\hat z\times\boldsymbol A(t)$, with $c_\eta=\pm1$ labeling the helicity $\eta$ of the drive. The drive displaces the electronic states along a circular momentum-space trajectory of radius $eA/\hbar$, and the response is controlled by the ratio of this radius to the Fermi momentum $k_F=\mu/(\hbar v_F)$. We therefore introduce
\begin{equation}\label{eq:xFdef}
    x_F\equiv\frac{k_F}{eA/\hbar},
\end{equation}
so that $x_F\gg1$ and $x_F\ll1$ define the weak- and strong-field regimes. Unless stated otherwise, we work in the slow-relaxation limit $\tau f\gg1$ and at low temperature, $k_BT\ll\mu,\ v_F eA$.

By the azimuthal and time-translation symmetry of the problem, the photocurrent can be decomposed as
\begin{equation}\label{eq:inplanedecomp}
    \boldsymbol{j}^{\eta}(t) =\bar j_z^{\eta }\hat z+  j_{P}^{\eta} \hat{\boldsymbol{e}}(t) + j_{N}^{\eta} \hat{z} \times \hat{\boldsymbol{e}}(t)
\end{equation}
[cf.~Eq.~(4) of the main text], where $\hat{\boldsymbol e}(t)\equiv{\boldsymbol E(t)}/{E}$ and the three coefficients $\bar j_z^{\eta}$, $j_P^{\eta}$, and $j_N^{\eta}$ are time independent in the steady state. Here $\bar j_z^\eta$ is the dc photocurrent along the propagation direction, while $j_P^\eta$ and $j_N^\eta$ are the in-plane components parallel and normal to the instantaneous electric field, which generate the screening response. It is useful to separate each in-plane component into the contribution of the group velocity and that of the anomalous velocity in Eq.~\eqref{eqa:phasevel}. As we show below, the anomalous velocity contributes only to the normal component, so that
\begin{equation}\label{eq:inplanedecomp_GH}
    j_P^\eta = j_{P,G}^\eta,\qquad
    j_N^\eta = j_{N,G}^\eta + j_N^H(\chi),
\end{equation}
where the subscript $G$ denotes the group-velocity (helicity-independent) contribution and $j_N^H(\chi)$ denotes the anomalous-velocity (helicity-dependent) contribution, which vanishes for $\chi=0$.

\subsection{Steady-state electronic distribution}\label{sec:ssocc}
We first derive an approximate analytic expression for the steady state occupation function  of the electronic bands, $g_\nu(\boldsymbol{k},t)$. {By explicitly solving the master equation [see Eq. (\ref{eq:rellinb})] \cite{PhysRevResearch.4.043060}, we obtain the solution} 
\begin{equation}
 \label{eq:ss}
    g_\nu(\boldsymbol{k},t) = \frac{1}{\tau} \int_{{-\infty}}^{t} ds \ e^{-(t-s)/\tau} f^{\text{eq}}_{\nu}(\boldsymbol{k},s),
\end{equation}
where $f^{\text{eq}}_{\nu}(\boldsymbol{k},t)$ defines the electronic occupation function of band $\nu$ in thermal equilibrium, and is given by
\begin{equation} \label{eq:feq}
    f^{\text{eq}}_{\nu}(\boldsymbol{k},t) = \frac{1}{e^{[\varepsilon_{\boldsymbol{k}\nu}(t)-\mu]/k_B T} + 1},
\end{equation}
where $\mu$ is the chemical potential, $k_B$ is the Boltzmann constant, and $T$ is the lattice temperature. 
As discussed in the main text, we focus on the slow-relaxation limit $\tau f\gg 1$, which is realized for {picosecond-scale relaxation times} and terahertz or higher frequency drives \cite{PhysRevB.95.085202,PhysRevB.95.041104}. In this limit, using Eq. (\ref{eq:ss}) and exploiting the time-periodicity of $f^{\text{eq}}_{\nu}(\boldsymbol{k},{t})$, we find that
\begin{equation} \label{eq:sslongtau}
    g_{\nu}(\boldsymbol{k},t) \approx  \frac{1}{T_{\text{dr}}} \int_0^{T_{\text{dr}}} du \  f^{\text{eq}}_{\nu} (\boldsymbol{k},u) \equiv \left\langle f^{\text{eq}}_{\nu} (\boldsymbol{k},u)\right\rangle_u,
\end{equation}
where $T_{\text{dr}} = 1/f$ is the period of the drive. {Thus, the steady-state distribution $g_{\nu}(\boldsymbol{k},t)$ approaches a distribution obtained from ``smearing'' the equilibrium distribution $f_\nu(\bk,t)$ along the circular trajectory $\boldsymbol{k}+e\boldsymbol{A}(t)/\hbar$ defined by the drive-induced vector potential.}

\subsubsection{Weak-field limit} \label{sec:occweakfield}
In the weak field limit $eA/\hbar \ll k_F$, the driving field weakly displaces the Fermi sphere, see Fig.~3(a) in the main text. Here we find an approximate expansion for the occupation function $g_1(\bk,t)$ up to order $A^2$ in the zero-temperature limit. 

For a single Weyl node with a positive chemical potential $\mu > 0$, we note that the instantaneous equilibrium distribution in the conduction band can be written as
\begin{equation}
    f^{\rm eq}_1(\bk,u)
    =
    f^{\rm eq}_1(\bk + e\boldsymbol{A}(u)/\hbar).
\end{equation}
where $f^{\rm eq}_1(\bk) = \theta(\mu - \varepsilon_{\bk \nu})$ in the zero-temperature limit, with $\theta$ denoting the Heaviside step function. 
Expanding in $eA/\hbar \ll k_F$ and averaging over $u$ gives
\begin{align}
    g_1(\bk,t)
    &=
    \left\langle f^{\rm eq}_1\!\left(\bk+e\bA(u)/\hbar\right)\right\rangle_u
    \nonumber\\
    &=
    f^{\rm eq}_1(\bk)
    + \frac{e}{\hbar} \sum_{i=x,y} \left\langle A_i(u)\right\rangle_u
       \partial_{k_i}f^{\rm eq}_1(\bk)
    +\frac{1}{2} \frac{e^2}{\hbar^2} \sum_{i,j=x,y} 
       \left\langle A_i(u)A_j(u)\right\rangle_u
       \partial_{k_i}\partial_{k_j}f^{\rm eq}_1(\bk)
    +\cdots .
    \label{eq:weak_occupation_expansion}
\end{align}
For a circularly polarized
driving field,
\begin{equation}
    \left\langle A_i(u)\right\rangle_u=0,
    \qquad
    \left\langle A_i(u)
    A_j(u)\right\rangle_u
    =
    \frac{A^2}{2}\delta_{ij}.
\end{equation}
Consequently,
\begin{equation}
    g_1(\bk,t)
    =
    f^{\rm eq}_1(\bk)
    +\frac{e^2A^2}{4\hbar^2}\nabla_{\bk_\perp}^2 f^{\rm eq}_1(\bk)
    +\cdots ,
    \label{eq:weak_occupation_general}
\end{equation}
where $\bk_{\perp} = (k_x,k_y)$. Thus, within the slow-relaxation approximation, the
drive-induced correction to the occupation in the weak field limit begins at order $A^2$.

\subsubsection{Strong-field limit} \label{sec:occstrongfield}
\begin{figure}
    \centering
    \includegraphics[width=0.7\linewidth]{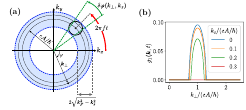}
    \caption{(a) Illustration of the characteristic in-plane momentum dynamics relevant for analyzing the photocurrent. Solid black circle indicates the edge of the instantaneous equilibrium Fermi volume of $H(\bk,t)$ as a function of in-plane momentum, $\bk_{\perp} = (k_x,k_y)$, for an arbitrary, fixed value of $k_z$. Black dot indicates the instantaneous location of the Weyl node, which is located at an azimuthal angle of $2\pi ft$ from the $k_x$ axis (red arrow). We approximate the instantaneous equilibrium occupation $f^{\text{eq}}_{1}(\boldsymbol{k},t)$ [Eq. (\ref{eq:esteqf})] to be given by $ 1$ if $\bk_\perp$ is located within the annulus segment (shaded blue) and between the two angles indicated by green lines, and $ 0$ otherwise. (For the choice of $\bk_{\perp}$ sketched in the figure, $\bk_{\perp}$ is positioned such that $f^{\text{eq}}_{1}(\boldsymbol{k},t) =0$.) Such an approximation is accurate in the limit where the Fermi momentum $k_F$ is much less than the vector potential $eA/\hbar$. (b) Occupation function $g_{1}(\boldsymbol{k},t)$ as a function of the in-plane momentum magnitude $k_{\perp} = \sqrt{k_x^2 +k_y^2}$ in the slow relaxation time limit $\tau f \gg 1$, plotted for several values of the $z$-momentum $k_z$.}
    \label{fig:coord}
\end{figure}

In the strong field limit $eA/\hbar \gg k_F$, the occupation function is smeared along a circular trajectory with a large radius $eA/\hbar \gg k_F$, see Fig.~3(b) in the main text. In this section, we specialize to the case of an isotropic Weyl node with no tilt $\chi = 0$. We revisit the case of $\chi \neq 0$ in Sec.~\ref{sec:occstrongfieldsmalltilt}. To find an approximate expression for the occupation function in this limit, we transform into the cylindrical coordinates $\boldsymbol{k} =  (k_{\perp} \cos\varphi, k_{\perp} \sin\varphi, k_z)$ and first find an approximate expression for the zero-temperature instantaneous equilibrium occupation function $f^{\text{eq}}_{1} (\boldsymbol{k},t) = \theta (k_F -|\boldsymbol{k}+e\boldsymbol{A}(t)/\hbar|)$, where $\theta$ is the Heaviside step function. In the regime $k_F \ll eA/\hbar$, it can be approximated by
\begin{equation} \label{eq:esteqf}
    f^{\text{eq}}_{1}(\boldsymbol{k},t) \approx \theta(\delta \varphi(k_{\perp},k_z)-\min_{n\in\mathbb{Z}} |\varphi - 2\pi f t + 2\pi n|) \theta(k_F^2 - k_z^2 - (k_{\perp} - eA/\hbar)^2),
\end{equation}
{where $\delta \varphi(k_{\perp}, k_z)$ measures the half-width of the azimuthal angle taken up by the Fermi volume at fixed $k_\perp$ and $k_z$ [see Fig.~\ref{fig:coord}(a)].    
The second step function in Eq. (\ref{eq:esteqf}) requires that the in-plane momentum magnitude $k_\perp$ of occupied states lies between the two dashed blue circles in Fig. \ref{fig:coord}(a), while the first step function in Eq. (\ref{eq:esteqf}) requires that the azimuthal angle $\varphi$ of the in-plane momentum lies between the two angles indicated by green dashed lines in Fig. \ref{fig:coord}(a). 
 It is convenient to denote the half-width of the angular window by $\delta \varphi(k_{\perp},k_z) = [\varphi_{+}(k_{\perp},k_z)-\varphi_-(k_{\perp},k_z)]/2$, where $\varphi_{\pm}(k_{\perp},k_z)$ are defined via the relation }
\begin{equation}\label{eq:dphicond}
    \hbar v_F |(k_{\perp} \cos[\varphi_{\pm}(k_{\perp},k_z)] - eA/\hbar, k_{\perp} \sin[\varphi_{\pm}(k_{\perp},k_z)], k_z)| = \hbar v_F k_F,
\end{equation}
which parameterizes the boundary of the Fermi volume. Expanding $\cos[\varphi_{\pm}(k_{\perp},k_z)]\approx1-[\varphi_{\pm}(k_{\perp},k_z)]^2/2$ in the limit $k_F \ll eA/\hbar$, solving for $\varphi_{\pm}(k_{\perp},k_z)$  yields 
\begin{equation} \label{eq:dphi}
    \delta \varphi(k_{\perp},k_z) \approx (\hbar/eA)\sqrt{k_F^2 - k_z^2 - ( k_{\perp} - eA/\hbar )^2}.
\end{equation}

The steady state occupation function [see Eq. (\ref{eq:sslongtau})] in the $\tau f \gg 1$ limit is the time-average of Eq. (\ref{eq:esteqf}), which is {time-independent} and given by
\begin{equation} \label{eq:occslowrelax}
    g_{1}(\boldsymbol{k},t) \approx \frac{1}{\pi} \left(\frac{\hbar}{eA}\right)\sqrt{k_F^2 - k_z^2 - (k_{\perp} - eA/\hbar)^2} \theta(k_F^2 - k_z^2 - (k_{\perp} - eA/\hbar)^2). 
\end{equation}
We plot $g_1(\boldsymbol{k},t)$ as a function of $k_{\perp}$ for several values of $k_z$ in Fig. \ref{fig:coord}(b).

\section{Photocurrent of an isolated, untilted Weyl node} \label{sec:untilted}
We first consider an isolated Weyl node without tilt, $\chi=0$, for which the Hamiltonian in Eq.~\eqref{eq:tiltedh} reduces to $H(\bk)=\xi\hbar v_F\bk\cdot\boldsymbol\sigma$. We compute the dc photocurrent along the propagation axis in Sec.~\ref{sec:zcurrsecuntilt}, and the oscillating in-plane photocurrent that generates the screening response in Secs.~\ref{sec:plasmon} and \ref{sec:weak_untilted}, in the strong- and weak-field regimes respectively. Throughout we work at low temperature, $k_BT\ll\mu$, and in the slow-relaxation limit $\tau f\gg1$.

\subsection{Out-of-plane photocurrent} \label{sec:zcurrsecuntilt}
For an untilted node, the out-of-plane $z$-current can be evaluated in closed form for arbitrary $x_F=k_F/(eA/\hbar)$. We analyze the contribution from each electronic band separately. To this end, we define $\bar{j}^z_{\eta} = \sum_{\nu} \bar j_{\nu,\eta}^z$,
where $\bar j_{\nu,\eta}^z$ is the time-averaged contribution to the current from the $\nu$-th band. Using the semiclassical equations of motion [Eqs.~\eqref{eqa:approx_current}-\eqref{eqa:phasevel}], we find 
\begin{equation} \label{eq:jnuexpr}
    \bar j_{\nu,\eta}^z = -\frac{e}{T_{\text{dr}}} \int_0^{T_{\text{dr}}} dt \int \frac{d^3 \boldsymbol{k}}{(2\pi)^3} \left[ \frac{ 1}{\hbar} \partial_{k_z}\varepsilon_{\boldsymbol{k}\nu}(t) - \left\{ e \boldsymbol{\Omega}_{\nu}\left(\boldsymbol{k} + \frac{e}{\hbar} \boldsymbol{A}(t)\right) \times \dfrac{\boldsymbol{E}(t)}{ \hbar} \right\} \cdot \hat{z}\right]  g_{\nu}(\boldsymbol{k},t).
\end{equation}
In untilted Weyl nodes, the occupation function $g_{\nu}(\boldsymbol{k},t)$ is symmetric upon inversion along $k_z$ ($k_z \to -k_z$), and $\partial_{k_z}\varepsilon_{\boldsymbol{k}\nu}(t)$ is an odd function of $k_z$. Therefore, the group velocity contribution vanishes. The remaining anomalous velocity is the origin of the topological current. For the untilted Weyl node considered here, the Berry curvature is given by   $\boldsymbol{\Omega}_{\nu}(\boldsymbol{k})=  {\xi} (-1)^{\nu} \boldsymbol{k} / 2 |\boldsymbol{k}|^3$. {Moreover using  that $\boldsymbol E(t)= -2\pi f c_\eta \hat z \times \boldsymbol A(t)$ for circularly polarized light}, we obtain  the photocurrent
\begin{equation}
\label{eq:jnufullint}
   {\bar j_{\nu,\eta}^z} = -{c_\eta \xi }(-1)^{\nu} \frac{e^2\omega}{2 \hbar}
   \frac{1}{T_{\rm dr}}\int_0^{T_{\rm dr}}\!dt \int \frac{d^3\boldsymbol{k}}{(2\pi)^3}\,
   \frac{[\boldsymbol{k}+e\boldsymbol{A}(t)/\hbar]\cdot\boldsymbol{A}(t)}
        {|{\boldsymbol k}+e{\boldsymbol A}(t)/\hbar|^{3}}\, g_{\nu}(\boldsymbol{k}) ,
\end{equation}
where we used that $g_\nu(\bk)$ is time independent in the slow-relaxation limit [see Eq.~\eqref{eq:sslongtau}].

For convenience, we next write Eq.~\eqref{eq:jnufullint} in a dimensionless form. Since the drive is circularly polarized, its direction $\hat{\boldsymbol A}(t)=\boldsymbol A(t)/A$ sweeps the unit circle at a uniform rate, and we parameterize it by an azimuthal angle $\alpha$,
\begin{equation}\label{eq:Ahat_alpha}
    \hat{\boldsymbol A}_\alpha\equiv(\cos\alpha,\sin\alpha,0),
    \qquad \hat{\boldsymbol A}(t)=\hat{\boldsymbol A}_{c_\eta\omega t},
\end{equation}
so that the time average in Eq.~\eqref{eq:jnufullint} amounts to an average over $\alpha$. Since $g_\nu(\bk)$ is invariant under rotations about $\hat z$, the integrand depends on $\alpha$ and on the azimuthal angle of $\bk$ only through their difference. We may therefore fix the drive vector potential along $\hat{\boldsymbol A}_0=\hat x$, and let $\phi$ denote the azimuthal angle of $\bk$ measured from the $x$-axis. Introducing, in addition to $x_F$ [see Eq.~\eqref{eq:xFdef}], the dimensionless momenta $x_{\perp}=k_{\perp}/(eA/\hbar)$ and $x_z=k_z/(eA/\hbar)$, where $k_{\perp}=|(k_x,k_y)|$, Eq.~\eqref{eq:jnufullint} becomes
\begin{equation} \label{eq:jnu_dimensionless}
    {\bar j_{\nu,\eta}^z} = -{c_\eta \xi }(-1)^{\nu}\frac{ef}{8\pi^2}\left(\frac{eA}{\hbar}\right)^{2}
    \int_{0}^{\infty}\! dx_{\perp}\,x_{\perp}\int_{-\infty}^{\infty}\! dx_z\int_0^{2\pi}\! d\phi\;
    \frac{x_{\perp}\cos\phi+1}{(x_{\perp}^2+2x_{\perp}\cos\phi+1+x_z^2)^{3/2}}\;
    \mathcal{G}_{\nu}(x_{\perp},x_z),
\end{equation}
where $\mathcal{G}_{\nu}(x_{\perp},x_z)\equiv g_{\nu}[(eA/\hbar)(x_{\perp},0,x_z)]$. We now compute the contributions from the $\nu=0$ and $\nu = 1$ bands separately, focusing on the case  $\mu \geq  0$; the results are easily generalized to  $\mu < 0$.

\subsubsection{Computing ${\bar j_{0,\eta}^z}$} \label{sec:compj0}
The valence band is fully occupied, so $\mathcal G_0(x_\perp,x_z)=1$ and Eq.~\eqref{eq:jnu_dimensionless} reduces to
\begin{equation}
    {\bar j_{0,\eta}^z} = -{c_\eta \xi }\frac{ef}{8\pi^2} \left( \frac{eA}{\hbar}\right)^2  \int_{0}^{\infty} dx_{\perp}\,x_{\perp} \int_0^{2\pi} d\phi \int_{-\infty}^{\infty} dx_z \frac{x_{\perp} \cos \phi + 1}{(x_{\perp}^2 + 2x_{\perp}\cos\phi + 1 + x_z^2)^{3/2}}.
\end{equation}
We first perform integration over $x_z$, which yields
\begin{equation}
    {\bar j_{0,\eta}^z}= -{c_\eta \xi }\frac{ ef}{8\pi^2} \left( \frac{eA}{\hbar}\right)^2  \int_{0}^{\infty} dx_{\perp} \int_0^{2\pi} d\phi \frac{2x_{\perp}(x_{\perp} \cos \phi + 1)}{x_{\perp}^2 + 1 + 2x_{\perp}\cos\phi}.
\end{equation}
Using the identity
\begin{equation}
    \int_0^{2\pi} d\phi \frac{\cos \phi + x_{\perp}^{-1}}{\cos \phi + (x_{\perp}^{-1} + x_{\perp})/2} = 4\pi \theta(1-x_{\perp}),
\end{equation}
it follows that ${\bar j_{0,\eta}^z} = {c_\eta \xi } j_{\text{top}}$, where
\begin{equation} \label{eq:jtop}
    j_{\text{top}} = -\frac{ef}{4\pi} \left(\frac{eA}{\hbar} \right)^2.
\end{equation}
{Thus, we have reproduced the quantized topological current density $j^z_{\rm top}$ quoted in the main text, which was obtained there using the charge pump picture [see the Thouless-pump discussion in the main text].}

\subsubsection{Computing ${j_{1,\eta}^z}$}
We next calculate the current ${j_{1,\eta}^z}$ generated by electronic states in the partially-filled $\nu = 1$ conduction band. 
Its occupation function is the cycle average of Eq.~\eqref{eq:sslongtau}, which at zero temperature reads
\begin{equation}\label{eq:g1_cycle}
    g_1(\bk)=\frac{1}{2\pi}\int_0^{2\pi}\!d\beta\;
    \theta\big(k_F-|\bk+(eA/\hbar)\hat{\boldsymbol A}_\beta|\big),
\end{equation}
i.e.\ the fraction of the drive cycle during which $\bk$ lies inside the instantaneous Fermi volume.
We return to the full expression of the $z$-current with dimensionful units in Eq.~\eqref{eq:jnufullint}, focusing on the conduction band $\nu=1$, and substitute Eq.~\eqref{eq:g1_cycle} for the occupation function. We further write  $\boldsymbol A(t)=A\hat{\boldsymbol A}_\alpha$, and replace the time average as an average over the azimuthal angle $\alpha$, i.e., $\frac{1}{T_{\mathrm{dr}}} \int_0^{T_{\mathrm{dr}}} dt \to \frac{1}{2\pi} \int_0^{2\pi} d\alpha$. Exchanging the integral over $\alpha$ and $\beta$ with the momentum integral, we obtain
\begin{equation}\label{eq:j1_double_average}
    {j_{1,\eta}^z}= c_\eta\xi\frac{e^2\omega A}{8\pi^2\hbar}
    \int_0^{2\pi}\!d\alpha\int_0^{2\pi}\!d\beta
    \int\frac{d^3\bk}{(2\pi)^3}\,
    \frac{[\bk+(eA/\hbar)\hat{\boldsymbol A}_\alpha]\cdot\hat{\boldsymbol A}_\alpha}
         {|\bk+(eA/\hbar)\hat{\boldsymbol A}_\alpha|^{3}}\;
    \theta\big(k_F-|\bk+(eA/\hbar)\hat{\boldsymbol A}_\beta|\big) .
\end{equation}

At fixed $\alpha$ and $\beta$, the step function $\theta$  can be made $\beta$-independent by shifting the momentum. Substituting
\begin{equation}
    \boldsymbol q=\frac{\hbar}{eA}\left[\bk+\frac{eA}{\hbar}\hat{\boldsymbol A}_\beta\right],
    \qquad d^3\bk=\left(\frac{eA}{\hbar}\right)^{3}d^3\boldsymbol q ,
\end{equation}
the step function becomes $\theta(x_F-|\boldsymbol q|)$, so the momentum integral runs over the ball $|\boldsymbol q|<x_F$, and we can replace $\bk+(eA/\hbar)\hat{\boldsymbol A}_\alpha=(eA/\hbar)(\boldsymbol q+\boldsymbol d_{\alpha\beta})$ with
\begin{equation}
    \boldsymbol d_{\alpha\beta}\equiv\hat{\boldsymbol A}_\alpha-\hat{\boldsymbol A}_\beta .
\end{equation}
Collecting the prefactors and using the definition of $j_{\mathrm{top}}$ [see Eq.~\eqref{eq:jtop}], we obtain
\begin{equation}\label{eq:j1_shifted}
    {j_{1,\eta}^z}=-c_\eta\xi\,\frac{j_{\rm top}}{8\pi^3}
    \int_0^{2\pi}\!d\alpha\int_0^{2\pi}\!d\beta
    \int_{|\boldsymbol q|<x_F}\! d^3\boldsymbol q\;
    \frac{(\boldsymbol q+\boldsymbol d_{\alpha\beta})\cdot\hat{\boldsymbol A}_\alpha}
         {|\boldsymbol q+\boldsymbol d_{\alpha\beta}|^{3}} .
\end{equation}
The integrand of Eq.~\eqref{eq:j1_shifted} is invariant under a simultaneous rotation of $\boldsymbol q$, $\hat{\boldsymbol A}_\alpha$ and $\hat{\boldsymbol A}_\beta$ about $\hat z$, so it depends on $\alpha$ and $\beta$ only through the difference $\beta-\alpha$. We may therefore set $\alpha=0$ without loss of generality, and the remaining integral over $\alpha$ trivial contributes a factor of $2\pi$, giving
\begin{equation}\label{eq:j1_beta_integral}
 -\frac{j_{1,\eta}^z}{c_\eta\xi j_{\rm top}}
 =\frac{1}{4\pi^2}\int_0^{2\pi}\!d\beta \left[\int_{|\boldsymbol q|<x_F}\!d^3\boldsymbol q\,
 \frac{(\boldsymbol q+\boldsymbol d_{\beta})}
 {|\boldsymbol q+\boldsymbol d_{\beta}|^3} \right]\cdot\hat{\boldsymbol A}_{0},
 \qquad
 \boldsymbol d_{\beta}\equiv\hat{\boldsymbol A}_0-\hat{\boldsymbol A}_{\beta}.
\end{equation}

The inner integral can be evaluated by Gauss' law, which gives
\begin{equation}\label{eq:gauss_ball}
 \int_{|\boldsymbol q|<x}d^3\boldsymbol q\,
 \frac{\boldsymbol q+\boldsymbol d}{|\boldsymbol q+\boldsymbol d|^3}
 =\frac{4\pi}{3}
 \begin{cases}
   \boldsymbol d, & d<x,\\[2pt]
   x^3\boldsymbol d/d^3, & d>x,
 \end{cases}
\end{equation}
To compute the dot product of Eq.~\eqref{eq:gauss_ball} with $\hat{\boldsymbol A}_0$, we note from Eq.~\eqref{eq:Ahat_alpha} that
\begin{equation}
    d_\beta=|\hat{\boldsymbol A}_0-\hat{\boldsymbol A}_\beta|=2\left|\sin(\beta/2)\right|,
    \qquad
    \hat{\boldsymbol A}_0\cdot\boldsymbol d_\beta=1-\cos\beta=\frac{d_\beta^2}{2}.
\end{equation}
Inserting this result into Eq.~\eqref{eq:j1_beta_integral} leaves a single integral over $\beta$,
\begin{equation}\label{eq:F_beta_integral}
    \mathcal F_1(x_F)\equiv-\frac{j_{1,\eta}^z}{c_\eta\xi j_{\rm top}}
    =\frac{1}{3\pi}\int_0^{2\pi}\!d\beta
    \left[\frac{d_\beta^{2}}{2}\,\theta(x_F-d_\beta)
    +\frac{x_F^{3}}{2\,d_\beta}\,\theta(d_\beta-x_F)\right].
\end{equation}
Substituting $u=\beta/2$, so that $d_\beta=2\sin u$ and $\int_0^{2\pi}d\beta=4\int_0^{\pi/2}du$, we find
\begin{equation}\label{eq:F_u_integral}
    \mathcal F_1(x_F)=\frac{2}{3\pi}\left[\int_0^{u_0}\!du\,4\sin^2u
    +\frac{x_F^{3}}{2}\int_{u_0}^{\pi/2}\frac{du}{\sin u}\right],
    \qquad u_0=\min\!\left[\arcsin(x_F/2),\,\tfrac{\pi}{2}\right].
\end{equation}
Both integrals are standard, $\int_0^{u_0}4\sin^2u\,du=2u_0-\sin 2u_0$ and $\int_{u_0}^{\pi/2}du/\sin u=\ln\cot(u_0/2)=\operatorname{arcosh}(2/x_F)$, which allows us to obtain the exact function
\begin{equation} \label{eq:exact_j1_crossover}
 {j_{1,\eta}^z}=-c_\eta\xi j_{\rm top}\,\mathcal F_1(x_F),\qquad
 \mathcal F_1(x)=
 \begin{cases}
 \displaystyle\frac{4\arcsin(x/2)-x\sqrt{4-x^2}+x^3\operatorname{arcosh}(2/x)}{3\pi}, & 0<x<2,\\[2ex]
 \displaystyle\frac{2}{3}, & x\geq 2.
 \end{cases}
\end{equation}

For $x_F\ll1$, Eq.~\eqref{eq:exact_j1_crossover} can be expanded as
\begin{equation}\label{eq:F_small_x}
 \mathcal F_1(x_F)=\frac{x_F^3}{3\pi}\left[\ln\!\left(\frac{4}{x_F}\right)+\frac{1}{3}\right]
 -\frac{x_F^5}{80\pi}+O(x_F^7),
\end{equation}
so the leading power is $x_F^3$, with a logarithmically varying prefactor. The total out-of-plane current of the node is given by $\bar j^z_{0,\eta}+j^z_{1,\eta}=c_\eta\xi j_{\rm top}[1-\mathcal F_1(x_F)]$, so the function $\mathcal F(x)$ entering Eq.~(9) of the main text is $\mathcal F(x)=1-\mathcal F_1(x)$. It decreases from $1$ at $x=0$ according to Eq.~\eqref{eq:F_small_x}, and saturates exactly at $1/3$ for $x_F\geq 2$, i.e., for $\mu\geq 2v_FeA$.

\begin{figure}
    \centering
    \includegraphics[width=0.4\linewidth]{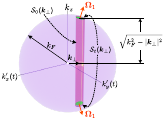}
    \caption{Illustration of the coordinate system used to calculate the saturation value of $j^z_{1,\eta}$ for $k_F \gg 2eA/\hbar$. Here, $\bk' = \bk + e\boldsymbol A(t)/\hbar$ is the shifted coordinate system, which fixes the WN at the origin. Purple sphere indicates the Fermi volume in the conduction band, which is centered at the WN (center white dot). Under illumination by circularly polarized light, populated electronic states with a given $\bk_{\perp} = (k_x, k_y)$ traces out a cylinder $\mathcal{S}_t(\bk_{\perp})$ (dark purple) in momentum space during a period of the drive. Due to the finite Fermi momentum $k_F$, finite Berry curvature $\boldsymbol \Omega_1$ penetrates the ``caps'' $\mathcal{S}_t(\bk_{\perp})$ (light green) of the cylinder located at $k_z = \pm \sqrt{k_F^2 - |\bk_{\perp}|^2}$. The Berry flux through the caps $\mathcal{S}_t(\bk_{\perp})$ produces a saturation value of $j^z_{1,\eta}$ in the limit $k_F \gg 2eA/\hbar$.}
    \label{fig:finitevolume}
\end{figure}

\paragraph*{Geometric interpretation of the saturation for $k_F \gg 2eA/\hbar$.}
Equation~(\ref{eq:exact_j1_crossover}) shows that ${j_{1,\eta}^z}$ saturates exactly once $k_F\geq2eA/\hbar$. For completeness, we give an alternative Berry-flux interpretation in the deep large-Fermi volume limit $k_F \gg 2eA/\hbar$.
To this end, we first consider the current produced by a single Fermionic chain of states with fixed in-plane momentum $\boldsymbol{k}_{\perp}$, given by 
\begin{equation}\label{eq:curidef}
    \mathcal{I}(\boldsymbol{k}_{\perp}) = -\frac{e}{T_{\text{dr}}} \int_0^{T_{\text{dr}}} dt\int_{-\infty}^{\infty} \frac{dk_z}{2\pi} \left[ \frac{ 1}{\hbar} \partial_{k_z}\varepsilon_{\boldsymbol{k}1}(t) - \left\{ e \boldsymbol{\Omega}_{1}\left(\boldsymbol{k} + \frac{e}{\hbar} \boldsymbol{A}(t)\right) \times \dfrac{\boldsymbol{E}(t)}{ \hbar} \right\} \cdot \hat{z}\right]  g_{1}(\boldsymbol{k},t).
\end{equation}
{This in-plane-momentum-resolved current determines $j_{1,\eta}^z$ via}
\begin{equation} \label{eq:int_i}
    {j_{1,\eta}^z} = \int \frac{d^2 \boldsymbol{k}_{\perp}}{(2\pi)^2} \mathcal{I}(\boldsymbol{k}_{\perp}).
\end{equation}
Noting again that the group velocity term in Eq. (\ref{eq:curidef}) vanishes by symmetry, it follows that
\begin{equation} 
    \mathcal{I}(\boldsymbol{k}_{\perp}) = \frac{e f}{2\pi} \int_{-\infty}^{\infty} dk_z \int_0^{T_{\text{dr}}} dt \ \frac{e}{\hbar} [\boldsymbol{\Omega}_1(\boldsymbol{k} + e \boldsymbol{A}(t)/\hbar) \times \boldsymbol{E}(t)] \cdot \hat{z} \ g_{1}(\boldsymbol{k},t).
\end{equation}
To further simplify the expression, we note that the magnetic vector potential $\boldsymbol{A}(t)$ is always perpendicular to the electric field $\boldsymbol{E}(t)$, and, therefore,
\begin{equation} \label{eq:ieq}
    \mathcal{I}(\boldsymbol{k}_{\perp})  = -{c_\eta}\frac{ef}{2\pi} \int_{-\infty}^{\infty} \int_0^{T_{\text{dr}}} \boldsymbol{\Omega}_1(\boldsymbol{k} + e \boldsymbol{A}(t)/\hbar) \cdot \left[ \frac{e}{\hbar} E \hat{\boldsymbol{A}}(t)  dt dk_z \right] g_{1}(\boldsymbol{k},t),
\end{equation}
where $E = |\boldsymbol{E}(t)|$ and $\hat{\boldsymbol{A}} = \boldsymbol{A}(t) / |\boldsymbol{A}(t)|$.
{We identify $\boldsymbol{\Omega}_1(\boldsymbol{k} + e \boldsymbol{A}(t)/\hbar)\cdot  ({e}/{\hbar}) E \hat{\boldsymbol{A}}(t)  dt dk_z$ [see Fig.~\ref{fig:finitevolume}] as the flux  of Berry curvature  through an infinitesimal surface area element for a cylinder $\mathcal S_0(\bk _\perp)$ of radius $eA/\hbar$ oriented along the $z$ axis and with center located at $\bk _\perp = (k_x,k_y)$. 
Next, we note that $g_1(\bk,t)\approx \theta(k_F-|\bk|)$ for $k_F\gg eA/\hbar$.
Thus $\mathcal{I}(\boldsymbol{k}_{\perp}) $ is given by the total Berry flux through the open surface (or ``ribbon'') defined by the segment of $\mathcal S_0(\bk_{\perp})$ with  $|k_z|<\sqrt{k_F^2-|\bk_\perp|^2} $:
\begin{equation}
     \mathcal{I}(\boldsymbol{k}_{\perp})  = -c_{\eta} \frac{ef}{2\pi}\int_{\mathcal S_0(\bk_\perp)} \boldsymbol \Omega _1\cdot d\boldsymbol S .
\end{equation}
To compute this, we consider the closed surface $\mathcal S(\bk _\perp)$  formed by closing the ends of $\mathcal S_0(\bk)$ by adding the disk-shaped ``caps'' $\mathcal S_t(\bk _\perp)$, consisting of points $\bk'=(k_x,k_y,\pm \sqrt{k_F^2-|\bk_\perp|^2})$, see Fig. \ref{fig:finitevolume} for an illustration. The surface integral over $\mathcal S_0(\bk_\perp)$ is given by the surface integral over $\mathcal S(\bk _\perp)$ minus the oriented surface integral over $\mathcal S_t(\bk_\perp)$: 
\begin{equation} \label{eq:i_eq_m}
    \mathcal{I}(\boldsymbol{k}_{\perp}) = -c_{\eta} \frac{ef}{2\pi} \left( \int_{\mathcal{S}(\bk_\perp)} \boldsymbol{\Omega}_1 \cdot d\boldsymbol{S} - \int_{\mathcal{S}_t (\bk_\perp)} \boldsymbol \Omega_1 \cdot d\boldsymbol{S} \right).
\end{equation}
The first integral gives the quantized Berry flux $-2\pi\xi$ when the closed surface encloses the Weyl node, and vanishes otherwise, following the charge-pump analysis in the main text.
To evaluate the second integral, we recall that  $\boldsymbol \Omega_1(\bk)=-\xi {\bk}/(2|\bk|^3)$ [see the definition of $\boldsymbol \Omega_\nu$ in Sec.~\ref{sec:zcurrsecuntilt}].
 When $k_F \gg  2eA/\hbar$, we can effectively set $k_z=\sqrt{k_F^2-|\bk_\perp|^2}$ on the surface  ${\mathcal{S}_t (\bk_\perp)} $, implying that the Berry flux density through ${\mathcal{S}_t (\bk_\perp)} $ is uniform and given by $-\xi   {\sqrt{k_F^2 - |\bk|^2}}/{k_F^3}$.   
Noting that ${\mathcal{S}_t (\bk_\perp)} $ has area $2\pi (eA/\hbar)^2$, this leads us to }
\begin{equation}
    \int_{\mathcal{S}_t{(\bk _\perp)} }\boldsymbol{\Omega}_1(\boldsymbol{k}) \cdot d\boldsymbol{S} \approx -\pi \xi \left(\frac{eA}{\hbar}\right)^2 \frac{\sqrt{k_F^2 - |\bk_{\perp}|^2}}{k_F^3}.
\end{equation}
Using Eq. (\ref{eq:int_i}) and Eq. (\ref{eq:i_eq_m}) to compute the total current produced all Fermionic chains in the conduction band, we find that
\begin{equation}
    {j_{1,\eta}^z} = c_{\eta} \xi \frac{2}{3} \frac{ef}{4\pi} \left(\frac{eA}{\hbar} \right)^2 \quad{\rm for}\quad k_F\gg 2eA/\hbar.
\end{equation}

\subsection{Strong-field in-plane response, $k_F \ll eA/\hbar$} \label{sec:plasmon}
We next compute the oscillating in-plane photocurrent in the strong-field regime $k_F\ll eA/\hbar$, in which the drive smears the Fermi volume into a thin torus of radius $eA/\hbar$ in momentum space [see Fig.~3(b) in the main text]. This is the regime responsible for the saturation of the polarization, and hence for the optical multistability.

{For concreteness, we set $t = 0$, noting that the results  for later times are related to this case through trivial rotation}. Using the semiclassical equations in {Eqs.~\eqref{eqa:approx_current} and \eqref{eqa:phasevel}, we find that the total current is given by
\begin{equation} \label{eq:inplanecurr}
    \boldsymbol{j}^\eta(t)= -e \sum_\nu \int \frac{d^3\boldsymbol{k}}{(2\pi)^3} \left[ \frac{ \nabla_{\boldsymbol{k}}}{\hbar} \varepsilon_{\boldsymbol{k}\nu}(t) - \left\{ e \boldsymbol{\Omega}_{\nu}\left(\boldsymbol{k} + \frac{e}{\hbar} \boldsymbol{A}(0)\right) \times \dfrac{\boldsymbol{E}(0)}{ \hbar} \right\} \right]_{\perp}  g_\nu(\boldsymbol{k},t),
\end{equation}
To evaluate the in-plane component of $\boldsymbol j^{\eta}(t)$, we first note that the anomalous velocity term does not contribute to the in-plane photocurrent for untilted Weyl nodes, because the in-plane component of the anomalous velocity reverses direction upon inversion along the $k_z$ axis ($k_z \to -k_z$), and the Fermi volume (and hence also $g_\nu(\bk,t)$)} is symmetric upon the same transformation. We also note that the $\nu = 0$ band does not contribute, because $g_0(\bk,t)=1$ and $\nabla_{\boldsymbol{k}} \varepsilon_{\boldsymbol{k}0}(0)$ is odd under momentum inversion about the Weyl node ($\bk + e\boldsymbol{A}(0)/\hbar \to -\bk + e\boldsymbol{A}(0)/\hbar$) and the Fermi volume is also symmetric about the same transformation. Thus,
\begin{equation}
    {j^\eta_{x,y}(t)=-e\int \frac{d^3\bk}{(2\pi)^{3}} \frac{1}{\hbar} \nabla_{k_x,k_y }\varepsilon_{\bk 1}(0)g_{1}(\bk,t)}.
\end{equation}

{We now evaluate the components $j^\eta_P$ and $j^\eta_N$ of the in-plane photocurrent parallel and normal to the instantaneous electric field, defined in Eq.~\eqref{eq:inplanedecomp}.
We identify $j_N^{\eta}=c_{\eta} j_x(0)$, $j_P^{\eta} =-c_{\eta} j_y(0)$. 
Also using the expression for the energies $\varepsilon_{\bk 1}(t)$, and exploiting the symmetry of $g_1(\bk,0)$ under $k_x\to-k_x$ to place the shifted node at $+eA/\hbar\,\hat x$, we find 
\begin{eqnarray}
    j^\eta_N  &=& c_\eta e v_F \int \frac{d^3\boldsymbol{k}}{(2\pi)^3} \frac{k_x - e A/\hbar}{\sqrt{[k_x - e A/\hbar ]^2 + k_y^2 + k_z^2}} g_1(\boldsymbol{k},0),
\\
     j^\eta_P &=& c_\eta e v_F  \int \frac{d^3\boldsymbol{k}}{(2\pi)^3} \frac{k_y}{\sqrt{[k_x - e A/\hbar ]^2 + k_y^2 + k_z^2}} g_1(\boldsymbol{k},0).
\end{eqnarray}}
To evaluate the integrals analytically, we use the strong-field approximation to the occupation function $g_1(\boldsymbol{k},0)$ given in Eq.~\eqref{eq:occslowrelax} and derived in Sec.~\ref{sec:occstrongfield}. {We exploit the rotational symmetry of $g_1(\boldsymbol{k},t)$ to express the occupation function in terms of the dimensionless cylindrical coordinates $x_\perp,x_z,\phi$, defined via $(k_x,k_y,k_z)=eA/\hbar (x_\perp \cos\phi,x_\perp\sin\phi,x_z)$, giving 
\begin{equation}
     \mathcal{G}(x_{\perp}, x_z, \phi) \equiv  g_{1}[eA/\hbar  (x_{\perp}\cos\phi ,x_{\perp}\sin\phi ,x_z),0].
\end{equation}}
In terms of this function, the in-plane components of the current read, for $\alpha = N,P$,
\begin{equation} \label{eq:weyltilteqint}
     j_{\alpha}^{\eta} = -c_\eta\frac{e v_F}{(2\pi)^3} \left( \frac{eA}{\hbar} \right)^3 \int_{0}^{\infty} dx_{\perp} \ x_{\perp} \int_0^{2\pi} d\phi \int_{-\infty}^{\infty} dx_z\frac{h_{\alpha}(\phi)}{\sqrt{x_{\perp}^2 + x_z^2 + 1 - 2 x_{\perp} \cos\phi}} \mathcal{G}(x_{\perp}, x_z, \phi),
\end{equation}
where $h_{N}(\phi) = 1-x_{\perp}\cos \phi$ and $h_{P}(\phi) = -x_{\perp}\sin \phi$. {Because $\mathcal{G}(x_{\perp}, x_z, \phi)$ is only nonzero when $|x_z|\leq  k_F/(eA/\hbar)$ and we assume $k_F \ll eA/\hbar $, we have  $x_z \ll 1$. As a result, we can approximate}
\begin{equation} \label{eq:intermedjnu}
     j_{\alpha}^{\eta} \approx  -c_{\eta} \frac{e v_F}{(2\pi)^3} \left( \frac{eA}{\hbar} \right)^3 \int_{0}^{\infty} dx_{\perp} \ x_{\perp} \int_0^{2\pi} d\phi \frac{h_{\alpha}(\phi)}{\sqrt{x_{\perp}^2 + 1 - 2 x_{\perp} \cos\phi}} \int_{-\infty}^{\infty} dx_z \ \mathcal{G}(x_{\perp}, x_z,\phi).
\end{equation}
{Converting the expression for $g_{1}(\boldsymbol{k},0)$ in Eq. (\ref{eq:occslowrelax})} to the cylindrical coordinates, we find
\begin{equation}\label{eq:occuntilt}
     \mathcal{G}(x_{\perp}, x_z, \phi) \approx \frac{1}{\pi} 
 \sqrt{x_F^2 - x_z^2 - (x_{\perp} - 1)^2} \theta(x_F^2 - x_z^2 - (x_{\perp} - 1)^2).
\end{equation}
We next perform the integral over $x_z$, finding 
\begin{equation} \label{eq:gint}
    \int_{-\sqrt{x_F^2 - (x_{\perp}-1)^2}}^{\sqrt{x_F^2 - (x_{\perp}-1)^2}} dx_z \ \mathcal{G}(x_{\perp}, x_z, \phi) = \frac{1}{2} [x_F^2 - (x_{\perp}-1)^2].
\end{equation}
Furthermore, in the $x_F \ll 1$ limit, we approximate
\begin{equation} \label{eq:current-xy}
    j_{\alpha}^{\eta} \approx -c_{\eta}  \frac{e v_F}{(2\pi)^3} \frac{1}{2\sqrt{2}} \left( \frac{eA}{\hbar} \right)^3 \left(\int_{1-x_F}^{1+x_F} dx \ x [x_F^2 - (x_{\perp}-1)^2]\right) \int_0^{2\pi} d\phi \frac{h_{\alpha}(\phi)}{\sqrt{1 - \cos\phi}}.
\end{equation}
Integration over $x$ and $\phi$ yields 
\begin{equation}
  j_{P}^{\eta} = 0\quad{\rm and}\quad   j_{N}^{\eta} \approx -c_{\eta} \frac{e v_F}{3\pi^3} x_F^3 \left( \frac{eA}{\hbar} \right)^3 \quad{\rm for}\quad k_F\ll eA/\hbar
\end{equation}
for the components parallel and perpendicular to the electric field $\boldsymbol E(t)$, respectively, where we have used that $h_{N}(\phi) \approx 1- \cos\phi$ throughout the integration range and $\int_0^{2\pi} d\phi \ \sqrt{1 - \cos\phi} = 4 \sqrt{2}$. This reproduces the strong-field current in Eq.~(6) of the main text.

\subsection{Weak-field in-plane response, $eA/\hbar \ll k_F$} \label{sec:weak_untilted}

We now consider the in-plane currents in the weak field limit $k_F \gg eA/\hbar$. Here, we apply a semiclassical analysis, which is valid in the quasi-adiabatic regime where $hf\ll \mu$. Our starting point is the occupation function in the slow relaxation and weak-field limit, which can be expanded as
\begin{equation}
    g_1(\boldsymbol k,t) \approx \theta(k_F-k)+O[(eA/\hbar k_F)^2],
    \label{eq:weak_untilted_occ}
\end{equation}
[see derivation in Sec.~\ref{sec:occweakfield}]. As in the strong-field case, the anomalous velocity, which is odd in $k_z$, gives no in-plane current for an untilted Weyl node, because the Fermi surface is symmetric about inversion along the $k_z$ axis, i.e., $k_z\to -k_z$. Thus the in-plane current arises entirely from the group velocity,
\begin{equation}
    \boldsymbol j_G^\eta(t)= -e v_F\int \frac{d^3\boldsymbol k}{(2\pi)^3}
    \frac{\boldsymbol k+e\boldsymbol A(t)/\hbar}{|\boldsymbol k+e\boldsymbol A(t)/\hbar|}\theta(k_F-k).
\end{equation}
Expanding the integrand for $eA/\hbar\ll k_F$ gives
\begin{equation} \label{eq:kpeaexp}
    \frac{\boldsymbol k+e\boldsymbol A(t)/\hbar}{|\boldsymbol k+e\boldsymbol A(t)/\hbar|}
    \approx \hat{\boldsymbol k}+\frac{e}{\hbar k}\left\{\boldsymbol A(t)-\hat{\boldsymbol k}[\hat{\boldsymbol k}\cdot \boldsymbol A(t)]\right\},
\end{equation}
where $\hat{\boldsymbol{k}}$ is the unit vector in the direction of $\boldsymbol{k}$. The first term in Eq.~\eqref{eq:kpeaexp} is odd under momentum inversion and integrates to zero. Using
\begin{equation}
    \int d\Omega_{\hat{\boldsymbol k}}\,\left(\delta_{ij}-\hat k_i\hat k_j\right)=\frac{8\pi}{3}\delta_{ij},
\end{equation}
where $\int d\Omega_{\hat{\boldsymbol k}}$ denotes integration over the solid angle of the unit vector $\hat{\boldsymbol k}=\boldsymbol k/k$, we obtain
\begin{equation}
    \boldsymbol j_G^\eta(t)\approx -\frac{e^2 v_F k_F^2}{6\pi^2\hbar}\boldsymbol A(t) \quad {\rm for}\quad k_F\gg eA/\hbar.
\end{equation}
Decomposing in terms of $j^\eta_{P/N,G}$ [see Eq.~\eqref{eq:inplanedecomp}], we thus have 
\begin{equation}
    j_{P,G}^\eta=0,\qquad c_\eta j_{N,G}^\eta\approx -\frac{e v_F k_F^2}{6\pi^2}\frac{eA}{\hbar} \quad {\rm for}\quad k_F\gg eA/\hbar.
    \label{eq:weak_weyl_group}
\end{equation}
This reproduces the weak-field Drude current in Eq.~(5) of the main text. 

We obtain the polarization from $\partial_t\boldsymbol P^\eta(t)=\boldsymbol j^\eta(t)$. Since the field direction rotates as $\partial_t\hat{\boldsymbol e}(t)=c_\eta\omega\,\hat z\times\hat{\boldsymbol e}(t)$, a polarization $\boldsymbol P^\eta(t)=P^\eta\hat{\boldsymbol e}(t)$ along the field generates precisely a current along $\hat z\times\hat{\boldsymbol e}(t)$. Comparing with Eq.~\eqref{eq:inplanedecomp} gives $P^\eta=c_\eta j_N^\eta/\omega$, and hence
\begin{equation}
    P_G^{\eta}\approx -\frac{e^2 v_F k_F^2}{6\pi^2\hbar\omega^2}E.
    \label{eq:weak_weyl_P}
\end{equation}
Thus, in the weak-field regime, the in-plane response is the ordinary linear Drude response of a Weyl cone (with the polarization antiparallel to the field), with stiffness
\begin{equation}
    \kappa_G\equiv -\omega^2\frac{P_G^{\eta}}{E}=\frac{e^2 v_F k_F^2}{6\pi^2\hbar}.
\end{equation}

\section{Consequences of finite longitudinal tilt} \label{sec:tilt}
We now consider a Weyl node with a finite tilt along the laser propagation axis, $\chi\neq0$. The tilt has two consequences. First, it leads to a group-velocity contribution to the out-of-plane current, producing a helicity-independent dc photocurrent along $z$, as derived in Sec.~\ref{sec:zcurrenttilt}. Second, by displacing the Fermi pocket along $k_z$, it allows the anomalous velocity to contribute a net in-plane current, producing a helicity-dependent component $j_N^H(\chi)$. We analyze these in-plane responses in the strong-field regime in Sec.~\ref{sec:inplanecurrtilt} and in the weak-field regime in Sec.~\ref{sec:weak_tilt}.

Before proceeding, we briefly discuss the geometry of the Fermi surface under finite tilt $\chi \neq 0$. We focus in particular on a type-I Weyl node, $|\chi|<1$, whose instantaneous conduction-band energy is given by
\begin{equation}
    \varepsilon_{\boldsymbol q,1}=\hbar v_F(|\boldsymbol q|+\chi q_z),
    \label{eq:tilted_disp}
\end{equation}
where $\boldsymbol q=\boldsymbol k+e\boldsymbol A(t)/\hbar$. At fixed chemical potential, the instantaneous Fermi volume is
\begin{equation}
    D_\chi=\{\boldsymbol q: |\boldsymbol q|+\chi q_z<k_F\}.
\end{equation}
Equivalently, the Fermi volume satisfies the condition
\begin{equation}
q_x^2+q_y^2+(1-\chi^2)\left(q_z+\frac{\chi k_F}{1-\chi^2}\right)^2<\frac{k_F^2}{1-\chi^2}.
\label{eq:tilted_ellipsoid}
\end{equation}
Thus the tilt changes the spherical Fermi volume into an ellipsoid shifted along $k_z$ by $-\chi k_F/(1-\chi^2)$, with volume
\begin{equation}
    V_F(\chi)=\frac{4\pi k_F^3}{3(1-\chi^2)^2}.
    \label{eq:tilted_volume}
\end{equation}
The shift along $k_z$ is the source of the helicity-dependent in-plane photocurrent, and the enlargement of the Fermi volume gives rise to an enhancement of the helicity-independent in-plane current.

\subsection{Out-of-plane photocurrent} \label{sec:zcurrenttilt}

When the Weyl node exhibits a tilt $\chi \neq 0$ along the $z$-axis, the topological photocurrent survives under weak anisotropy $\chi \ll 1$, as demonstrated numerically in Fig.~5 of the main text. However, the anisotropy allows for the group velocity term in the semiclassical equations of motion [see Eqs.~(2)--(3) of the main text] to contribute a nonzero, laser helicity-independent photocurrent [see Fig.~5(a) in the main text], given by
\begin{equation} \label{eq:genjg}
    \bar j^z_G =  -\frac{e}{T_{\text{dr}}} \int_0^{T_{\text{dr}}} dt \int \frac{d^3 \boldsymbol{k}}{(2\pi)^3} \left[ \frac{1}{\hbar} \frac{\partial\varepsilon_{\boldsymbol{k}1}(t)}{\partial k_z} \right]   g_{1}(\boldsymbol{k},t).
\end{equation}
In the slow relaxation limit $\tau f \gg 1$, we can rewrite 
\begin{equation}
    g_{1}(\boldsymbol{k},t) \approx  \frac{1}{T_{\text{dr}}} \int_0^{T_{\text{dr}}} dt' \  f^{\text{eq}}_{1} (\boldsymbol{k},t')
\end{equation}
[see Eq. (\ref{eq:sslongtau})] and express the instantaneous equilibrium occupation function in terms of the instantaneous eigenenergies
\begin{equation}
    f^{\text{eq}}_{1} (\boldsymbol{k},t) = \theta(\mu-\varepsilon_{\boldsymbol{k}1}(t)).
\end{equation}

To provide intuition for the emergence of the helicity-independent photocurrent, we note that the tilt produces an asymmetric $f^{\text{eq}}_{1} (\boldsymbol{k},t)$ that preferentially populates electronic states with $z$-momentum $k_z$ antiparallel to the direction of anisotropy, i.e., $\text{sign}(k_z)=-\text{sign}(\chi)$. The steady state occupation $g_{1}(\boldsymbol{k},t)$, which time-averages the instantaneous electronic occupation, therefore acquires larger values for electronic states {whose $z$-momentum has sign}
$-\text{sign}(\chi)$; the tilt term in the dispersion endows this imbalanced population with a net group velocity along $\text{sign}(\chi)\hat z$ and hence, for carriers of charge $-e$, a net electric current along $-\text{sign}(\chi)\hat z$.

\begin{figure}
    \centering
    \includegraphics[width=0.25\linewidth]{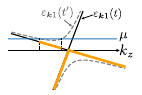}
    \caption{Origin of the laser helicity-independent photocurrent in a tilted WN. In the limit $\tau f \gg 1 $, the steady state occupation of bands at any given time $t$ is the time average of the instantaneous equilibrium occupation function $\theta(\mu-\varepsilon_{\boldsymbol{k}1}(t'))$ at all other times $t'$. The black solid and dashed gray curves represent the electronic bands at representative times $t$ and $t'$, respectively. The instantaneous equilibrium distribution at time $t'$ (orange) preferentially fills states with $\partial \varepsilon_{\boldsymbol{k}1}(t) / \partial k_z < 0$, producing a net helicity-independent current.}
    \label{fig:helindcurr}
\end{figure}

To estimate the helicity-independent photocurrent, we rewrite Eq. (\ref{eq:genjg}) as
\begin{equation} \label{eq:jglag}
    \bar j_{G}^z = -ef \int \frac{dk_x dk_y}{(2\pi)^2}  \int_0^{T_{\text{dr}}} dt'  \mathcal{J}_z(k_x, k_y, t, t'),
\end{equation}
where
\begin{equation} \label{eq:inthelind}
    \mathcal{J}_z(k_x, k_y, t, t') \equiv \int \frac{dk_z}{2\pi} \frac{1}{\hbar} \frac{\partial \varepsilon_{\boldsymbol{k}1}(t)}{\partial k_z} \theta(\mu-\varepsilon_{\boldsymbol{k}1}(t')).
\end{equation}
As discussed previously, the slow relaxation time $\tau$ gives rise to a lagging of the Fermi volume along the trajectory of the Weyl node traced out by the vector potential. This is captured in Eq. (\ref{eq:jglag}) because the current at time $t$ is given by the average of the occupation-weighted group velocity $\mathcal{J}_z(k_x, k_y, t, t')$ set by the instantaneous thermal equilibrium distribution at all other times $t'$. 
Crucially, $\mathcal{J}_z(k_x, k_y, t, t')$ is nonzero with sign given by $\text{sign}(\chi)$ only when $\chi \neq 0$ and $t \neq t'$, as illustrated in Fig. \ref{fig:helindcurr}, which is the origin of the helicity-independent photocurrent.
When $\chi < 0$, the electronic momenta with nonzero instantaneous equilibrium occupation $\theta(\mu-\varepsilon_{\boldsymbol{k}1}(t'))$ (thick orange line) at previous time $t' < t$ preferentially populates electronic states with negative $z$-group velocity ${\partial \varepsilon_{\boldsymbol{k}1}(t)}/{\partial k_z}<0$ at time $t$, allowing the integral in Eq. (\ref{eq:inthelind}) to attain a nonzero negative value.
Eq. (\ref{eq:inthelind}) can be solved by integration by parts, yielding
\begin{equation} \label{eq:intpart}
    \mathcal{J}_z(k_x, k_y, t, t') = \frac{1}{2\pi \hbar} \left[ \varepsilon_{\boldsymbol{k}1}(t)|_{k_z=k_z^+(t')} - \varepsilon_{\boldsymbol{k}1}(t)|_{k_z = k_z^-(t')} \right],
\end{equation}
where $\varepsilon_{\bk 1}(t')|_{k_z = k_z^{\pm}(t')} = \mu$. To compute the values of $k_z^{\pm}(t')$, we consider the instantaneous eigenenergy at time $t'$, given by
\begin{equation}
    \varepsilon_{\boldsymbol{k}1}(t') = \hbar v_F |\boldsymbol{k}+e\boldsymbol{A}(t')/\hbar| + \hbar v_F \chi k_z.
\end{equation}
Defining $k_F \equiv \mu /(\hbar v_F)$ and solving for the instantaneous Fermi surface $\varepsilon_{\boldsymbol{k}1}(t') = \hbar v_F k_F$ gives rise to two roots
\begin{equation}
    k_z^{\pm}(t') \equiv \frac{\chi k_F \mp \sqrt{(\chi^2-1)[{\tilde \bk}_{\perp}(t')]^2+k_F^2}}{\chi^2-1},
\end{equation}
where 
\begin{equation}
    {\tilde \bk}_{\perp}(t') \equiv (k_x + eA_x(t')/\hbar, k_y+eA_y(t')/\hbar)
\end{equation}
denotes the in-plane momentum {shifted by the drive-induced vector potential }via minimal coupling. 

Therefore, using Eq.~(\ref{eq:intpart}), we find
\begin{align}
    \mathcal{J}_z(k_x,k_y,t,t')
    ={}&\frac{v_F}{\pi}
    \sqrt{k_F^2-\left|\tilde{\boldsymbol{k}}_\perp(t')\right|^2}
    \left[
        1-
        \frac{k_F}{\sqrt{
            k_F^2
            +\left|\tilde{\boldsymbol{k}}_\perp(t)\right|^2
            -\left|\tilde{\boldsymbol{k}}_\perp(t')\right|^2
        }}
    \right]
    \nonumber\\
    &\times
    \theta\!\left(k_F^2-\left|\tilde{\boldsymbol{k}}_\perp(t')\right|^2\right)
    \chi+O(\chi^3).
    \label{eq:largejzexprn}
\end{align} 

To expand in the weak-field limit $eA/\hbar\ll k_F$, we define
$\boldsymbol{p}_\perp\equiv\tilde{\boldsymbol{k}}_\perp(t')$ and
$\Delta\tilde{\boldsymbol{k}}_\perp\equiv
\tilde{\boldsymbol{k}}_\perp(t)-\tilde{\boldsymbol{k}}_\perp(t')
=(e/\hbar)[\boldsymbol{A}(t)-\boldsymbol{A}(t')]$.
We also define
\begin{equation}
    \left|\tilde{\boldsymbol{k}}_\perp(t)\right|^2
    -\left|\tilde{\boldsymbol{k}}_\perp(t')\right|^2
    =2\boldsymbol{p}_\perp\cdot\Delta\tilde{\boldsymbol{k}}_\perp
    +\left|\Delta\tilde{\boldsymbol{k}}_\perp\right|^2
    \equiv X.
\end{equation}
The term in brackets in Eq.~\eqref{eq:largejzexprn} can then be expanded as
\begin{align}
    1-\frac{k_F}{\sqrt{k_F^2+X}}
    & =1-\left(1+\frac{X}{k_F^2}\right)^{-1/2}
    \nonumber\\
    & \approx
    \frac{X}{2k_F^2}-\frac{3X^2}{8k_F^4}
    \nonumber\\
    & =
    \frac{\boldsymbol{p}_\perp\cdot
    \Delta\tilde{\boldsymbol{k}}_\perp}{k_F^2}
    +\frac{\left|\Delta\tilde{\boldsymbol{k}}_\perp\right|^2}{2k_F^2}
    -\frac{3\left(
    \boldsymbol{p}_\perp\cdot\Delta\tilde{\boldsymbol{k}}_\perp
    \right)^2}{2k_F^4},
\end{align}
where terms of third and higher order in
$\Delta\tilde{\boldsymbol{k}}_\perp$ have been discarded. Substitution into
Eq.~\eqref{eq:largejzexprn} gives
\begin{align}
    \mathcal{J}_z(k_x,k_y,t,t')\approx{}&
    \frac{v_F}{\pi}\sqrt{k_F^2-p_\perp^2}
    \left[
        \frac{\boldsymbol{p}_\perp\cdot
        \Delta\tilde{\boldsymbol{k}}_\perp}{k_F^2}
        +\frac{\left|\Delta\tilde{\boldsymbol{k}}_\perp\right|^2}{2k_F^2}
        -\frac{3\left(
        \boldsymbol{p}_\perp\cdot\Delta\tilde{\boldsymbol{k}}_\perp
        \right)^2}{2k_F^4}
    \right]
    \theta(k_F^2-p_\perp^2)\chi.
\end{align}
The term linear in $\Delta\tilde{\boldsymbol{k}}_\perp$ vanishes upon
angular integration, while
\begin{equation}
    \int_0^{2\pi}d\varphi\,
    \left(
        \boldsymbol{p}_\perp\cdot\Delta\tilde{\boldsymbol{k}}_\perp
    \right)^2
    =\pi p_\perp^2
    \left|\Delta\tilde{\boldsymbol{k}}_\perp\right|^2.
\end{equation}
Consequently,
\begin{equation}
    \int_0^{2\pi}d\varphi\,\mathcal{J}_z
    \approx
    v_F\chi\sqrt{k_F^2-p_\perp^2}
    \left|\Delta\tilde{\boldsymbol{k}}_\perp\right|^2
    \left(
        \frac{1}{k_F^2}-\frac{3p_\perp^2}{2k_F^4}
    \right)
    \theta(k_F^2-p_\perp^2).
\end{equation}
For circular polarization, the remaining time average is
\begin{equation}
    \frac{1}{T_{\mathrm{dr}}} \int_0^{T_{\mathrm{dr}}}dt'\,
    \left|\Delta\tilde{\boldsymbol{k}}_\perp\right|^2
    =2\left(\frac{eA}{\hbar}\right)^2.
\end{equation}
Substituting these results into Eq.~(\ref{eq:jglag}) gives
\begin{align}
    \bar{j}_G^z
    ={}&-\frac{e v_F\chi}{\pi^2}
    \left(\frac{eA}{\hbar}\right)^2
    \int_0^{k_F}dp_\perp\,p_\perp\sqrt{k_F^2-p_\perp^2}
    \left(
        \frac{1}{2k_F^2}-\frac{3p_\perp^2}{4k_F^4}
    \right).
\end{align}
Using the radial integrals
\begin{equation}
    \int_0^{k_F}dp_\perp\,p_\perp\sqrt{k_F^2-p_\perp^2}
    =\frac{k_F^3}{3},
    \qquad
    \int_0^{k_F}dp_\perp\,p_\perp^3\sqrt{k_F^2-p_\perp^2}
    =\frac{2k_F^5}{15},
\end{equation}
we have
\begin{equation}
    \int_0^{k_F}dp_\perp\,p_\perp\sqrt{k_F^2-p_\perp^2}
    \left(
        \frac{1}{2k_F^2}-\frac{3p_\perp^2}{4k_F^4}
    \right)
    =\frac{k_F}{6}-\frac{k_F}{10}
    =\frac{k_F}{15}.
\end{equation}
Using $\mu=\hbar v_F k_F$, we finally obtain
\begin{align}
    \bar{j}_G^z
    ={}&-\frac{e\mu}{15\pi^2\hbar}
    \left(\frac{eA}{\hbar}\right)^2\chi
    \qquad\text{for}\qquad k_F\gg\frac{eA}{\hbar},
\end{align}
which reproduces the tilt current $\bar{j}_{\mathrm{tilt}}^z$ in Eq.~(10) of the main text.

\subsection{Strong-field in-plane response, $k_F \ll eA/\hbar$} \label{sec:inplanecurrtilt}
We now discuss the effect of a tilt on the in-plane photocurrent in the strong-field limit $k_F\ll eA/\hbar$. In Sec.~\ref{sec:occstrongfieldsmalltilt}, we compute the effect of a weak tilt ($\chi\ll1$) on the occupation function. In Sec.~\ref{sec:topinplane}, we use this result to obtain the helicity-dependent anomalous-velocity current $j_N^H(\chi)$ to leading order in $\chi$. This term gives rise to the helicity-selective plasmonic response discussed in the main text [see Figs.~1(c) and 4(b) in the main text]. In Sec.~\ref{sec:modinplanechi}, we compute the effect of the tilt on the helicity-independent in-plane current, finding that it is enhanced at $O(\chi^2)$ due to the enlargement of the Fermi volume. Finally, in Sec.~\ref{sec:strongP}, we combine the two contributions to compute the polarization function in the strong field limit.

\subsubsection{Strong-field occupation function at small tilt} \label{sec:occstrongfieldsmalltilt}

The tilt along the $z$-axis reshapes the instantaneous equilibrium occupation function $f^{\text{eq}}_1(\boldsymbol{k},t)$ in the conduction band of the Weyl node, which in turn modifies the steady state occupation function $g_1(\bk,t)$. To recalculate $g_1(\bk, t)$, we revisit the derivation in Sec. \ref{sec:occstrongfield}, using the approximate expression for the occupation function in the slow relaxation limit $\tau f \gg 1$ given by
\begin{equation} \label{eq:g1int}
    g_{1}(\boldsymbol{k},t) \approx  \frac{1}{T_{\text{dr}}} \int_0^{T_{\text{dr}}} du \  f^{\text{eq}}_{1} (\boldsymbol{k},u).
\end{equation}
In the presence of a tilt, we again have 
\begin{equation}\label{eq:feqtilt}
f^{\mathrm{eq}}_{1}(\boldsymbol{k},t)
\approx
\theta\!\left(
\delta\varphi_\chi(k_\perp,k_z)
-\min_{n\in\mathbb Z}
|\varphi-2\pi f t+2\pi n|
\right)
\theta\!\left(
(k_F-\chi k_z)^2-k_z^2
-(k_\perp-eA/\hbar)^2
\right).
\end{equation}
[see Eq. (\ref{eq:esteqf})], but with a modified $\delta \varphi_\chi (k_\perp,k_z)$ defined by $\delta \varphi_\chi (k_\perp,k_z)=[\varphi_{\chi,+}(k_\perp,k_z)-\varphi_{\chi,-}(k_\perp,k_z)]/2$, where
\begin{equation}
\hbar v_F
\left|
\left(
k_\perp\cos[\varphi_{\chi,\pm}(k_\perp,k_z)]-eA/\hbar,\,
k_\perp\sin[\varphi_{\chi,\pm}(k_\perp,k_z)],\,
k_z
\right)
\right|
+\hbar v_F\chi k_z
=
\hbar v_F k_F .
\end{equation}
parameterizes the edge of the Fermi volume [see Eq. (\ref{eq:dphicond})] which may be rewritten in the form
\begin{equation}
    (1+\chi^2) k_F^2 = k_{\perp}^2 - 2\frac{eA}{\hbar} k_{\perp} \cos[\varphi_{\chi,\pm}(k_{\perp},k_z) ]+ \frac{e^2A^2}{\hbar^2} + (\sqrt{1-\chi^2} k_z +\chi k_F)^2 + 2(1-\sqrt{1-\chi^2}) \chi k_F k_z.
\end{equation}
Expanding $\cos[\varphi_{\chi,\pm}(k_{\perp},k_z) ]\approx1-[\varphi_{\chi,\pm}(k_{\perp},k_z)]^2/2$ in the limit $k_F \ll eA/\hbar$ and evaluating $\delta \varphi_\chi (k_\perp,k_z)=[\varphi_{\chi,+}(k_\perp,k_z)-\varphi_{\chi,-}(k_\perp,k_z)]/2$, we find that
\begin{equation}
    [\delta \varphi_{\chi} (k_{\perp}, k_z)]^2 \approx (\hbar/eA)^2 [{k_F^2 - (k_z +\chi k_F)^2  - (k_{\perp} - eA/\hbar)^2]},
\end{equation}
where we have discarded terms of $O(\chi^2)$ or higher.

Using Eq. (\ref{eq:g1int}), we time-average the equilibrium occupation function, Eq. (\ref{eq:feqtilt}), to find
\begin{equation} 
    g_{1}(\boldsymbol{k},t) \approx \frac{1}{\pi} \left(\frac{\hbar}{eA}\right)\sqrt{k_F^2 - (k_z+\chi k_F)^2 -(k_{\perp}-eA/\hbar)^2} \theta(k_F^2 - (k_z+\chi k_F)^2 -(k_{\perp}-eA/\hbar)^2).
    \label{eqa:g1tilt}
\end{equation}
Note that $  g_{1}(\boldsymbol{k},t) $ is time-independent.
We see that the primary effect of the tilt in the limit of weak anisotropy $\chi \ll 1$ is a shift of the Fermi volume along the $k_z$ axis by $-\chi k_F$. Expressed in the dimensionless cylindrical coordinates $x_\perp,x_z,\phi$ introduced in Sec.~\ref{sec:plasmon}, the occupation function reads
\begin{equation} \label{eq:occtilt}
    \mathcal{G}(x_{\perp}, x_z,\phi) \approx \frac{1}{\pi} \sqrt{x_F^2 - (x_z+\delta)^2 -(x_{\perp}-1)^2} \theta(x_F^2 - (x_z+\delta)^2 -(x_{\perp}-1)^2),
\end{equation}
where $\delta \equiv \chi x_F$ measures the shift of the Fermi volume in units of $eA/\hbar$. 
We use Eq.~\eqref{eq:occtilt} in the two subsections that follow.

\subsubsection{Anomalous-velocity contribution $j_{N}^H$ to leading order in the tilt}\label{sec:topinplane}
Due to the asymmetry of the electronic occupation function along the $k_z$-axis, and the combination of rotation and time-translation symmetry of the problem, the anomalous velocity in Eq. (\ref{eq:inplanecurr}) contributes a photocurrent given by
\begin{equation} 
    \boldsymbol{j}_{\perp}^a = e \int \frac{d^3\boldsymbol{k}}{(2\pi)^3} \left[ e \boldsymbol{\Omega}_{1}\left(\boldsymbol{k} + \frac{e}{\hbar} \boldsymbol{A}(0)\right) \cdot \hat{z}\right] \left[ \hat{z}\times \dfrac{\boldsymbol{E}(0)}{ \hbar}  \right]  g_1(\boldsymbol{k},0).
\end{equation}
The current flows perpendicular to $\boldsymbol{E}(0)$ and is given by $\boldsymbol{j}_{\perp}^a = j_{N}^H(\chi) \hat{z} \times \hat{\boldsymbol e}(0)$, where
\begin{align}
    j_{N}^H(\chi) = -\xi \frac{e^2}{2\hbar} E \int \frac{d^3\boldsymbol{k}}{(2\pi)^3} \frac{k_z}{\{[k_x + e A/\hbar ]^2 + k_y^2 + k_z^2\}^{3/2}} g_1(\boldsymbol{k},0).
\end{align}
To evaluate the integral, we again transform to the dimensionless coordinates given by $x_F = k_F/(eA/\hbar)$, $x_{\perp} = k_{\perp} /(eA/\hbar)$, and $x_z = k_z /(eA/\hbar)$, {such that     $\mathcal{G}(x_{\perp}, x_z, \phi) \equiv  g_{1}[eA/\hbar  (x_{\perp}\cos\phi ,x_{\perp}\sin\phi ,x_z),0]$}.
This leads to
\begin{equation}
    j_{N}^H(\chi)  = -\xi \frac{ef}{8\pi^2} \left( \frac{eA}{\hbar} \right)^2 \int_0^{\infty} dx_{\perp} \ x_{\perp} \int_0^{2\pi} d\phi \int_{-\infty}^{\infty} dx_z \frac{x_z}{(x_{\perp}^2 + 1 +2x_{\perp}\cos\phi +x_z^2)^{3/2}} \mathcal{G}(x_{\perp}, x_z,\phi).
\end{equation}
Inserting the occupation function of Eq.~\eqref{eq:occtilt}, with $\delta=\chi x_F$, this leads to 
\begin{equation}
    j_{N}^H(\chi) \approx  -\xi  \frac{ef}{8\pi^3} \left( \frac{eA}{\hbar} \right)^2 \int_{-x_F 
- \delta}^{x_F - \delta} dx_z \int_{1-\sqrt{x_F^2 - (x_z + \delta)^2}}^{1+\sqrt{x_F^2 - (x_z + \delta)^2}} dx_{\perp} \ x_{\perp} \int_0^{2\pi} d\phi  \frac{\sqrt{x_F^2 - (x_z+\delta)^2 - (x_{\perp} - 1)^2}}{[x_{\perp}^2 + 1 + 2x_{\perp}\cos\phi +x_z^2]^{3/2}} x_z.
\end{equation}
We first perform integration over $\phi$ by expanding the integrand around $\phi \sim \pi$, where the integrand is largest
\begin{equation}
    \int_0^{2\pi} d\phi  \frac{1}{[x_{\perp}^2 + 1 + 2x_{\perp}\cos\phi +x_z^2]^{3/2}} \approx \frac{1}{x_{\perp}^{1/2}} \frac{2}{x_z^2   + \delta x_{\perp}^2},
\end{equation}
where $\delta x_{\perp} \equiv x_{\perp} - 1$. Therefore,
\begin{equation}
    j_{N}^H(\chi) \approx-\xi \frac{ef}{4\pi^3} \left( \frac{eA}{\hbar} \right)^2  \int_{-x_F 
- \delta}^{x_F - \delta} dx_z \int_{-\sqrt{x_F^2 - (x_z + \delta)^2}}^{\sqrt{x_F^2 - (x_z + \delta)^2}} d\delta x \frac{\sqrt{x_F^2 - (x_z+\delta)^2 - \delta x_{\perp}^2}}{x_z^2  + \delta x_{\perp}^2} x_z.
\end{equation}
Transforming to the shifted coordinates $x_z'=x_z+\delta$, we find
\begin{equation}
    j_{N}^H(\chi) \approx -\xi \frac{ef}{4\pi^3} \left( \frac{eA}{\hbar} \right)^2  \int_{-x_F}^{x_F} dx_z' \int_{-\sqrt{x_F^2 - (x_z')^2}}^{\sqrt{x_F^2 - (x_z')^2}} d\delta x \frac{\sqrt{x_F^2 - (x_z')^2 - \delta x_{\perp}^2}}{(x_z' -\delta)^2  + \delta x_{\perp}^2} (x_z'-\delta).
\end{equation}
We now use the integral identity
\begin{equation}
    \int_{-a}^{a} dx \frac{\sqrt{a^2-x^2}}{b^2+x^2} = \pi \left[ \sqrt{1+\frac{a^2}{b^2}}-1 \right]
\end{equation}
to obtain 
\begin{equation}
    \int_{-\sqrt{x_F^2 - (x_z')^2}}^{\sqrt{x_F^2 - (x_z')^2}} d\delta x \frac{\sqrt{x_F^2 - (x_z')^2 - \delta x_{\perp}^2}}{(x_z' -\delta)^2  + \delta x_{\perp}^2} = \pi \sqrt{1+\frac{x_F^2 - (x_z')^2}{(x_z'-\delta)^2}} - \pi.
\end{equation}
Therefore,
\begin{equation}
    j_{N}^H(\chi) = -\xi \frac{ef}{4\pi^2} \left( \frac{eA}{\hbar} \right)^2  \int_{-x_F}^{x_F} dx_z' \left[ \text{sign}(x_z'-\delta)  \sqrt{x_F^2 - (x_z')^2 +(x_z'-\delta)^2} - (x_z'-\delta)\right].
\end{equation}
Finally, upon integrating over $x_z'$ and expanding in $\chi$, we find, using $x_F\,eA/\hbar=k_F$,
\begin{equation}
    j_{N}^H(\chi) \approx \xi \frac{ef}{4\pi^2}\,k_F^2\,\chi \quad{\rm for}\quad k_F\ll eA/\hbar.
    \label{eq:jnh_strong}
\end{equation}

\subsubsection{Group-velocity contribution $j^{\eta}_{N,G}$ under arbitrary tilt}\label{sec:modinplanechi}
We next consider the in-plane current produced by the group velocity of the electrons. In contrast to the anomalous-velocity contribution of Sec.~\ref{sec:topinplane}, this contribution can be obtained for an arbitrary type-I tilt, without expanding in $\chi$.

The tilt adds a constant velocity $v_F\chi\hat z$ to the group velocity [see Eq.~\eqref{eq:tilted_disp}], which has no in-plane component. The in-plane group-velocity current therefore takes the same form as for the untilted node [see Eq.~\eqref{eq:inplanecurr}]. Setting $t=0$ as in Sec.~\ref{sec:plasmon} and taking $\boldsymbol A(0)=A\hat x$, it reads
\begin{equation} \label{eq:jng}
    \boldsymbol j_{G}^{\eta}(0) = -e v_F \int \frac{d^3\boldsymbol{k}}{(2\pi)^3}\,
    \frac{\boldsymbol{k} + (eA/\hbar)\hat x}{|\boldsymbol{k} + (eA/\hbar)\hat x|}\; g_1(\boldsymbol{k},0),
\end{equation}
now evaluated with the occupation function of the tilted node. The components normal and parallel to the field follow from $j_{N,G}^{\eta}=c_{\eta}j_{x,G}(0)$ and $j_{P,G}^{\eta}=c_{\eta}j_{y,G}(0)$ [see Sec.~\ref{sec:plasmon}].

To evaluate the current, we substitute $g_1(\boldsymbol{k},0)= T_{\rm dr}^{-1}\int_0^{T_{\rm dr}}\! du\, f^{\text{eq}}_{1}(\boldsymbol{k},u)$ [see Eq.~\eqref{eq:g1int}] into Eq.~\eqref{eq:jng} and exchange the time average with the momentum integral:
\begin{equation}\label{eq:jng_timeavg}
    \boldsymbol j_{G}^{\eta}(0) = -\frac{e v_F}{T_{\rm dr}}\int_0^{T_{\rm dr}}\! du \int \frac{d^3\boldsymbol{k}}{(2\pi)^3}\,
    \frac{\boldsymbol{k} + e\boldsymbol A(0)/\hbar}{|\boldsymbol{k} + e\boldsymbol A(0)/\hbar|}\, f^{\text{eq}}_{1}(\boldsymbol{k},u).
\end{equation}
At zero temperature, the instantaneous equilibrium occupation of the tilted node is given by $f^{\text{eq}}_{1}(\boldsymbol{k},u)=\theta(k_F-|\boldsymbol{k}+e\boldsymbol A(u)/\hbar|-\chi k_z)$. For convenience, we exchange $\bk$ for $\boldsymbol q=\boldsymbol{k}+e\boldsymbol A(u)/\hbar$, i.e., the momentum measured from the instantaneous position of the Weyl node. Under this substitution, the domain of the momentum integration becomes the static tilted Fermi volume $D_\chi$, see Eq.~\eqref{eq:tilted_ellipsoid}, while $\boldsymbol{k}+e\boldsymbol A(0)/\hbar=\boldsymbol q+e[\boldsymbol A(0)-\boldsymbol A(u)]/\hbar$ and thus
\begin{equation}
\boldsymbol j_G^\eta(0)=-\frac{e v_F}{T_{\rm dr}}\int_0^{T_{\rm dr}}\!du\int_{D_\chi}\frac{d^3\boldsymbol q}{(2\pi)^3}
\frac{\boldsymbol q+e[\boldsymbol A(0)-\boldsymbol A(u)]/\hbar}{|\boldsymbol q+e[\boldsymbol A(0)-\boldsymbol A(u)]/\hbar|}.
\label{eq:strong_swap_integral}
\end{equation}
We can make simplifications to the integral in the strong field limit. Here $e[\boldsymbol A(0)-\boldsymbol A(u)]/\hbar$ has a magnitude of order $eA/\hbar$ throughout most of the integration domain $0 \leq u \leq T_{\mathrm{dr}}$. The momentum $\boldsymbol q$ is instead bounded by the size of the Fermi volume: since $|q_z|\leq|\boldsymbol q|$, the condition $|\boldsymbol q|+\chi q_z<k_F$ defining $D_\chi$ implies $|\boldsymbol q|<k_F/(1-|\chi|)$. In the strong-field regime $eA/\hbar\gg k_F/(1-|\chi|)$, where the drive displacement is large compared with the Fermi volume, we may therefore neglect $\boldsymbol q$ in the integrand of Eq.~\eqref{eq:strong_swap_integral}. With the integrand now momentum-independent, we have
\begin{equation}
    \boldsymbol j_G^\eta(0)\approx -\frac{e v_F k_F^3}{6\pi^2(1-\chi^2)^2}\;
    \frac{1}{T_{\rm dr}}\int_0^{T_{\rm dr}}\!du\;
    \frac{\boldsymbol A(0)-\boldsymbol A(u)}{|\boldsymbol A(0)-\boldsymbol A(u)|} ,
\end{equation}
where the momentum integral has been replaced with the Fermi volume  $V_F(\chi)/(2\pi)^3$ [see Eq.~\eqref{eq:tilted_volume}]. Writing $\varphi=\omega u$, the integral identity
\begin{equation}
\frac{1}{2\pi}\int_0^{2\pi}d\varphi\,
\frac{1-\cos\varphi}{\sqrt{(1-\cos\varphi)^2+\sin^2\varphi}}=\frac{2}{\pi},
\end{equation}
gives
\begin{equation}
    j_{P,G}^\eta(\chi)=0,\qquad
    c_\eta j_{N,G}^\eta(\chi)\approx
    -\frac{e v_F k_F^3}{3\pi^3}\frac{1}{(1-\chi^2)^2}
    \quad{\rm for}\quad \frac{eA}{\hbar}\gg\frac{k_F}{1-|\chi|}.
    \label{eq:strong_group_finite_chi}
\end{equation}
Here, the normal component is enhanced by a factor of $(1-\chi^2)^{-2}$ relative to the isotropic case. This accounts for the tilt-dependence of the helicity-independent induced field shown in Fig.~4(a) of the main text.

\subsubsection{Combined strong-field polarization}\label{sec:strongP}
Combining the group-velocity result of Eq.~\eqref{eq:strong_group_finite_chi} with the anomalous-velocity result of Eq.~\eqref{eq:jnh_strong}, and using $P^\eta=c_\eta j_N^\eta/\omega$ [see Eq.~\eqref{eq:inplanedecomp_GH} and Sec.~\ref{sec:weak_untilted}], the strong-field polarization of a single tilted Weyl node is
\begin{equation}
    P_\eta^{\rm strong}\approx
    -\frac{e v_F k_F^3}{3\pi^3\omega}\frac{1}{(1-\chi^2)^2}
    +\xi c_\eta\frac{e\chi k_F^2}{8\pi^3}
    \quad {\rm for}\quad k_F\ll eA/\hbar.
    \label{eq:strong_P_summary}
\end{equation}

\subsection{Weak-field in-plane response, $eA/\hbar \ll k_F$} \label{sec:weak_tilt}
We now consider the in-plane response of the tilted Weyl node in the weak-field regime $eA/\hbar\ll k_F$, where the drive only weakly displaces the tilted Fermi pocket. As for the untilted node (Sec.~\ref{sec:weak_untilted}), the helicity-independent group-velocity contribution produces a Drude response. In Sec.~\ref{sec:weakgroupchi} we evaluate it for an arbitrary type-I tilt, and find that the tilt renormalizes the Drude weight by the factor $F_\chi$ quoted in the main text. The tilt also displaces the Fermi pocket along $k_z$, which allows the in-plane anomalous velocity contribute an in-plane photocurrent. In Sec.~\ref{sec:weakanomalous} we evaluate the resulting helicity-dependent current to first order in $\chi$. In Sec.~\ref{sec:weakP} we combine the two contributions to compute the weak-field polarization of the tilted node.

\subsubsection{Group-velocity response for an arbitrary type-I tilt}\label{sec:weakgroupchi}

In the weak-field regime, the occupation function of the tilted node is, to linear order in the vector potential, simply its equilibrium occupation,
\begin{equation}
    g_1(\boldsymbol k,t)= \theta(k_F-k-\chi k_z)+O[(eA/\hbar k_F)^2],
    \label{eq:weak_tilt_occ_exact}
\end{equation}
where $k=|\boldsymbol k|$. This follows exactly as in the untilted case [see Eq.~\eqref{eq:weak_untilted_occ} and Sec.~\ref{sec:occweakfield}]: the contribution linear in $\boldsymbol A(t)$ averages to zero over the drive cycle. The support of Eq.~\eqref{eq:weak_tilt_occ_exact} is the tilted Fermi pocket $D_\chi$, defined in Eq.~\eqref{eq:tilted_ellipsoid}.

The tilt velocity $v_F\chi\hat z$ has no in-plane component, so the in-plane group-velocity current can be obtained by integrating
\begin{equation}
    \boldsymbol j_G^\eta(t)= -e v_F\int_{D_\chi} \frac{d^3\boldsymbol k}{(2\pi)^3}
    \frac{\boldsymbol k+e\boldsymbol A(t)/\hbar}{|\boldsymbol k+e\boldsymbol A(t)/\hbar|}
    \label{eq:weak_tilt_jG}
\end{equation}
and projecting the result onto the $x$-$y$ plane. Due to the azimuthal symmetry of the problem, the currents at any two points in time are related to one another by a trivial rotation about the $z$ axis, so we may take $\boldsymbol A(t)=A\hat x$ without loss of generality, and extract the normal and perpendicular components of the in-plane current via $j^\eta_{N,G}=c_\eta j^\eta_{x,G}$ and $j^\eta_{P,G}=c_\eta j^\eta_{y,G}$ [see Sec.~\ref{sec:plasmon}]. Expanding the integrand of Eq.~\eqref{eq:weak_tilt_jG} to linear order in the vector potential [cf.~Eq.~\eqref{eq:kpeaexp}] gives
\begin{equation}\label{eq:weak_tilt_expansion}
    \frac{\boldsymbol k+(eA/\hbar)\hat x}{|\boldsymbol k+(eA/\hbar)\hat x|}
    \approx \hat{\boldsymbol k}+\frac{eA}{\hbar k}\left(\hat x-\hat k_x\hat{\boldsymbol k}\right),
\end{equation}
whose in-plane components are given by $\hat k_x+(eA/\hbar k)(1-\hat k_x^2)$ along $\hat x$, and $\hat k_y-(eA/\hbar k)\,\hat k_x\hat k_y$ along $\hat y$. Since the pocket $D_\chi$ depends on $\boldsymbol k$ only through $k_\perp$ and $k_z$ [see Eq.~\eqref{eq:tilted_ellipsoid}], it is invariant under $k_x\to-k_x$ and under $k_y\to-k_y$ separately. The term $\hat k_x$ is odd under the inversion operation $k_x \to -k_x$, and the terms $\hat k_y$ and $\hat k_x\hat k_y$ are odd under $k_y \to -k_y$. Thus these three terms must vanish after integration over $D_\chi$, leaving only 
\begin{equation}
    j^\eta_{x,G}=-e v_F \frac{eA}{\hbar}\int_{D_\chi}\frac{d^3\boldsymbol k}{(2\pi)^3}
    \frac{1-\hat k_x^2}{k}+O[(eA/\hbar)^3].
\end{equation}
In spherical coordinates, the volume of $D_\chi$ is bounded by $k<k_F/(1+\chi u)$, where $u=\cos\theta=k_z/k$. The integral therefore becomes
\begin{equation}
    \int_{D_\chi}\frac{d^3\boldsymbol k}{(2\pi)^3}
    \frac{1-\hat k_x^2}{k}
    =\frac{k_F^2}{16\pi^2}\int_{-1}^{1}du\,\frac{1+u^2}{(1+\chi u)^2}.
\end{equation}
Thus
\begin{equation}
    j_{P,G}^\eta(\chi)=0,\qquad
    c_\eta j_{N,G}^\eta(\chi)\approx
    -\frac{e v_F k_F^2}{6\pi^2}\frac{eA}{\hbar}F_\chi
    \quad {\rm for}\quad k_F\gg eA/\hbar,
    \label{eq:weak_group_finite_chi}
\end{equation}
where
\begin{equation}
    F_\chi=\frac{3}{8}\int_{-1}^{1}du\,\frac{1+u^2}{(1+\chi u)^2}
    =\frac{3\left[2\chi-(1-\chi^2)\ln\left(\frac{1+\chi}{1-\chi}\right)\right]}{4\chi^3(1-\chi^2)}.
    \label{eq:Fw_chi}
\end{equation}
This is the Drude-weight renormalization factor $F_\chi$ quoted in the main text. In the limit $\chi\to0$,
\begin{equation}
    F_\chi=1+\frac{6}{5}\chi^2+\frac{9}{7}\chi^4+O(\chi^6).
\end{equation}
Equation~\eqref{eq:weak_group_finite_chi} shows that, as in the strong-field regime [Eq.~\eqref{eq:strong_group_finite_chi}], the helicity-independent in-plane response is unchanged to first order in the tilt, but can nevertheless acquire corrections at finite $\chi$ due to the enlargement of the Fermi volume.

\subsubsection{Anomalous-velocity contribution to leading order in the tilt}\label{sec:weakanomalous}
We next evaluate the anomalous-velocity contribution to the in-plane current in the weak-field limit, up to first order in $\chi$. The tilt enters the occupation function in the weak field limit, see Eq.~\eqref{eq:weak_tilt_occ_exact}, only through the shift $\chi k_z$ in the argument of the step function. Using $\theta(x-\epsilon)=\theta(x)-\epsilon\,\delta(x)+O(\epsilon^2)$, which follows from $\theta'(x)=\delta(x)$, with $x=k_F-k$ and $\epsilon=\chi k_z$, we obtain
\begin{equation}
    g_1(\boldsymbol k,t)\approx \theta(k_F-k)-\chi k_z\delta(k_F-k)
    \label{eq:weak_tilt_occ}
\end{equation}
to first order in $\chi$.
The second term in Eq.~\eqref{eq:weak_tilt_occ} is odd in $k_z$ and therefore does not modify the group-velocity contribution to the in-plane current. However, it allows the in-plane anomalous velocity to contribute a finite in-plane current under integration along the $k_z$ axis. Using the Berry curvature $\boldsymbol\Omega_1(\boldsymbol k)=-\xi \boldsymbol k/(2k^3)$ [see Sec.~\ref{sec:zcurrsecuntilt}], the anomalous in-plane current is given by
\begin{equation}
    \boldsymbol j_H(t)=e\int \frac{d^3\boldsymbol k}{(2\pi)^3}
    \left[e\boldsymbol\Omega_1(\boldsymbol k)\times \frac{\boldsymbol E(t)}{\hbar}\right]g_1(\boldsymbol k,t).
\end{equation}
Due to azimuthal symmetry, only the $z$ component of the Berry curvature contributes. Thus, from Eq.~\eqref{eq:weak_tilt_occ}, we obtain 
\begin{equation}
    \int \frac{d^3\boldsymbol k}{(2\pi)^3}\Omega_{1,z}(\boldsymbol k)g_1(\boldsymbol k,t)
    =\xi\chi\int \frac{d^3\boldsymbol k}{(2\pi)^3}\frac{k_z^2}{2k^3}\delta(k_F-k)
    =\xi\chi\frac{k_F}{12\pi^2}.
\end{equation}
Therefore,
\begin{equation}
    j_N^H(\chi)\approx \xi\chi\frac{e^2 k_F}{12\pi^2\hbar}E\quad {\rm for}\quad k_F\gg eA/\hbar.
    \label{eq:weak_tilt_current}
\end{equation}

\subsubsection{Combined weak-field polarization}\label{sec:weakP}
Combining Eqs.~\eqref{eq:weak_group_finite_chi} and \eqref{eq:weak_tilt_current}, and again using $P^\eta=c_\eta j_N^\eta/\omega$, the weak-field polarization of a single tilted Weyl node is given by
\begin{equation}
    P_\eta^{\rm weak}\approx
    -\left[
    \frac{e^2 v_F k_F^2}{6\pi^2\hbar\omega^2}F_\chi
    -\xi c_\eta\chi\frac{e^2 k_F}{12\pi^2\hbar\omega}
    \right]E.
    \label{eq:weak_tilt_P}
\end{equation}

\section{Stability Analysis from the Lindblad Master Equation} \label{sec:stability}
\newcommand{\brho}{\boldsymbol{\rho}}

In this section, we present in full detail the stability analysis of the self-consistent solutions to the plasmonic response, as summarized in the End Matter. The analysis below considers Weyl semimetals with azimuthal symmetry about the laser propagation axis, for which the self-consistent state contains only the fundamental harmonic. Models with broken azimuthal symmetry, whose self-consistent steady states contain higher harmonics, are treated separately in the End Matter.

We begin by deriving the eigenvalue problem that governs perturbations of a self-consistent state. {Throughout this section, we will consider a single self-consistent solution; we let %
$\rho_0(\bk,t)$ denote the (time-periodic) evolution of the electronic density matrix in this solution [resulting from Eq.~(11) of the main text], let  $\boldsymbol P_0(t)$ denote the electric polarization generated by the surface charge oscillation due to in-plane currents, and let $\boldsymbol A_0(t)$ denote the resulting screened internal magnetic vector potential as obtained by the Maxwell relations [Eq.~(1) of the main text]. All these quantities  are  time periodic with period $T_{\rm dr}=2\pi/\omega$. We also assume a perfectly closed circuit along the $z$-direction which fixes $A_{0,z} (t) = 0$.  We encompass this self-consistent solution in the vector $\boldsymbol v_0(t)=(\boldsymbol A_0(t),\boldsymbol P_0(t),\{\rho_0(\bk,t)\}_{\bk})$. Relevant for us will also be the Hamiltonian experienced by the electrons in this self-consistent solution, which we denote %
\begin{equation}
    h(\boldsymbol k,t)\equiv H[\boldsymbol k+
e\boldsymbol A_0(t)/\hbar],
\end{equation}
where $H(\bk)$ is the undriven single-particle Hamiltonian.

We infer the stability of the steady state above by obtaining the response of the system to adding an arbitrary weak perturbation to the self-consistent solution, $\bv_0(t)\to \bv_0(t)+\delta \bv(t)$, with $\delta\boldsymbol v(t)=(\delta\boldsymbol A(t),\delta\boldsymbol P(t),\{\delta\rho(\bk,t)\}_{\bk})^T$. Specifically, we shall  obtain the evolution of $\delta\boldsymbol v(t)$ to linear order in the perturbation. %
To infer the evolution of $\delta\boldsymbol v(t)$, and hence determine the stability of the self-consistent solution $\bv_0(t)$, we first linearize the Maxwell and Lindblad equations that govern the evolution of $\bv(t)$ under this perturbation [see Eq.~(1) in the main text and Eqs.~(11) and (12) in the End Matter]. This yields a linear system of equations of the form
\begin{equation}
 \partial_t\delta\boldsymbol v=\mathcal L(t)\delta\boldsymbol v,
 \qquad
 \mathcal L(t)
 =\begin{pmatrix}
  0&\kappa I_2&0\\
  0&0&\mathcal C\\
  \mathcal B&0&\mathcal K
 \end{pmatrix},
 \qquad \kappa\equiv\frac{N_\parallel}{\epsilon_0}.
 \label{eq:stab_jacobian}
\end{equation}
The first row of $\mathcal{L}(t)$ follows from $\boldsymbol E_{\rm ind}=-\kappa\boldsymbol P$ and $\boldsymbol E=-\partial_t\boldsymbol A$, while the second row relates the current to the density-matrix perturbation $\delta \rho(\bk,t)$, with 
\begin{equation}
 \mathcal C[\{\delta\rho(\bk')\}_{\bk'}](t)
 =-\frac{e}{\hbar}\int\frac{d^3\boldsymbol k}{(2\pi)^3}
 \operatorname{Tr}\!\left[\delta\rho(\bk)
 \nabla_{\boldsymbol k}h(\bk,t)\right].
 \label{eq:c(t)}
\end{equation}
The third row follows from the correction to the right-hand side of the Lindblad master equation [Eq.~(11) of the main text] arising from the perturbations of $\rho(\bk,t)$ and $\boldsymbol A(t)$.  %
Specifically, through minimal coupling, the perturbation $\delta\boldsymbol A(t)$ shifts both the Hamiltonian and the instantaneous equilibrium density matrix, $h\to h+(e/\hbar)\delta\boldsymbol A(t)\cdot\nabla_{\boldsymbol k_\perp}h$ and $\rho^{{\rm eq}}_0\to\rho^{{\rm eq}}_0+(e/\hbar)\delta\boldsymbol A(t)\cdot\nabla_{\boldsymbol k_\perp}\rho^{{\rm eq}}_0$.  Inserting both variations into the master equation and retaining terms linear in $\delta\rho$ and $\delta\boldsymbol A$ gives
$\partial_t\delta\rho(\bk,t)=\mathcal B(t)[\delta\boldsymbol A](\bk)
+\mathcal K(t)[\delta\rho](\bk)$, where
\begin{equation}
 \mathcal K[\{\delta\rho(\bk')\}_{\bk'}](\bk,t)
 =\frac{i}{\hbar}[\delta\rho(\bk),h(\bk,t)]
 -\frac{\delta\rho(\bk)}{\tau},
 \label{eq:k(t)}
\end{equation}
and
\begin{equation}
 \mathcal B[\delta\boldsymbol A](\bk,t)
 =\frac{ie}{\hbar^2}\left[\rho_0(\bk,t),\,
 \delta \boldsymbol A\cdot\nabla_{\boldsymbol k}h(\bk,t)\right]
 +\frac{e}{\hbar\tau}\,\delta\boldsymbol A\cdot
 \nabla_{\boldsymbol k}\rho^{\rm eq}_0(\bk,t).
 \label{eq:b(t)}
\end{equation}

Since $h(\bk,t)$, $\rho_0(\bk,t)$, and $\rho^{\rm eq}_0(\bk,t)$ are periodic, so is $\mathcal L$: $\mathcal L(t+T_{\rm dr})=\mathcal L(t)$. Floquet's theorem therefore allows the solutions to the above equations to be written as a linear combination of solutions of the form  $\delta\boldsymbol v_\lambda(t)=e^{\lambda t}\boldsymbol u_\lambda(t)$, with $\boldsymbol u_\lambda(t) = (\delta \boldsymbol{A}_{\lambda}(t), \delta \boldsymbol{P}_{\lambda}(t), \{\delta \rho_{\lambda}(\bk,t) \}_{\bk})$ periodic. Here, the decay or growth rate is given by $\operatorname{Re}\lambda$, and unstable modes are identified with $\operatorname{Re} \lambda > 0$. 

We can therefore conveniently recast Eq.~\eqref{eq:stab_jacobian} as a stationary problem, using the Sambe space formulation~\cite{PhysRevA.7.2203}. To this end, we expand the three periodic components of $\boldsymbol u_\lambda(t)$ in harmonics of the drive,
\begin{align}
    \delta\boldsymbol A_{\lambda}(t)&=\sum_m \delta \tilde{\boldsymbol{A}}_{\lambda,m} e^{im\omega t} \\
    \delta\boldsymbol P_{\lambda}(t)&=\sum_m \delta \tilde{\boldsymbol P}_{\lambda,m} e^{im\omega t} \\
    \delta\rho_{\lambda}(\bk,t)&=\sum_m\delta\tilde{\rho}_{\lambda,m}(\bk)\,e^{im\omega t},
\end{align}
and stack their coefficients into
\begin{align}
    \delta \tilde{\boldsymbol{{A}}}_{\lambda} &=(\ldots,\delta \tilde{\boldsymbol A}_{\lambda,-1},\delta \tilde{\boldsymbol A}_{\lambda,0},\delta \tilde{\boldsymbol A}_{\lambda,1},\ldots)^T \\
    \delta \tilde{\boldsymbol P}_{\lambda}&=(\ldots,\delta \tilde{\boldsymbol P}_{\lambda,-1},\delta \tilde{\boldsymbol P}_{\lambda,0},\delta \tilde{\boldsymbol P}_{\lambda,1},\ldots)^T \\
    \delta\tilde{\boldsymbol\rho}_{\lambda} &=(\ldots,\{\delta\tilde \rho_{\lambda,-1} (\bk)\}_{\boldsymbol k},\{\delta\tilde \rho_{\lambda,0}(\bk)\}_{\boldsymbol k},\{\delta\tilde \rho_{\lambda,1}(\bk)\}_{\boldsymbol k},\ldots)^T.
\end{align}
Defining $\tilde{\boldsymbol{u}}_{\lambda} = (\delta \tilde{\boldsymbol{{A}}}_{\lambda},  \delta \tilde{\boldsymbol P}_{\lambda},\delta\tilde{\boldsymbol\rho}_{\lambda})$, we substitute the Fourier expansions into Eq.~\eqref{eq:stab_jacobian}, which gives the explicit Floquet eigenvalue problem
\begin{equation}
 \begin{gathered}
  \tilde{\mathcal{M}}_{\rm F}\tilde{\boldsymbol u}_{\lambda}=\lambda\tilde{\boldsymbol u}_{\lambda},
  \qquad
  \tilde{\mathcal M}_{\rm F}=
  \begin{pmatrix}
   -i\Omega&\kappa I&0\\
   0&-i\Omega&\tilde{\mathcal C}\\
  \tilde{ \mathcal B}&0&-i\Omega+\tilde{\mathcal K}
  \end{pmatrix},
 \end{gathered}
 \label{eq:stab_floquet_reduction}
\end{equation}
where $\Omega_{nm}\equiv n\omega\delta_{nm}$ and
\begin{align}
    (\tilde{\mathcal C}\delta\tilde{\boldsymbol\rho}_{\lambda})_n
 &=-\frac{e}{\hbar}\int\frac{d^3\boldsymbol k}{(2\pi)^3}
 \operatorname{Tr}\!\left\{\sum_m\delta\tilde \rho_{\lambda,m}(\boldsymbol k )
 \nabla_{\boldsymbol k}\tilde h_{n-m}(\bk)\right\} \\
 (\tilde{\mathcal K}\delta\tilde{\boldsymbol\rho}_{\lambda})_{ n}
 &=\frac{i}{\hbar}\sum_m
 [\delta\tilde \rho_{ \lambda,m}(\boldsymbol k),\tilde h_{n-m}(\bk)]
 -\frac{\delta\tilde \rho_{\lambda, n}(\boldsymbol{k})}{\tau}\\
 (\tilde{\mathcal B} \delta \tilde{\boldsymbol{A}}_{\lambda})_{n}
 &=\frac{i}{\hbar}\sum_\ell \left[{\tilde{\rho}_{0,\ell}}(\bk),
 \frac{e}{\hbar}\sum_m
\delta \tilde{\boldsymbol A}_{\lambda,m}\cdot\nabla_{\boldsymbol k}
\tilde h_{n-\ell-m}(\bk)\right]
 +\frac{e}{\hbar \tau}\sum_m \delta \tilde{\boldsymbol A}_{\lambda,m}\cdot
\nabla_{\boldsymbol k} \tilde \rho^{{\rm eq}}_{0,n-m}(\bk),
\end{align}
where $\tilde\rho_{0,m}(\bk)$, $\tilde{h}_m(\bk)$, and $\tilde\rho^{\rm eq}_{0,m}(\bk)$ are defined via 
\begin{align}
    \rho_0(\bk,t) &= \sum_m \tilde\rho_{0,m}(\bk) e^{im\omega t}\\
    h(\bk,t) &= \sum_m \tilde{h}_m(\bk) e^{im\omega t}\\
    \rho^{\rm eq}_0(\bk,t)\equiv\rho^{\rm eq}(\bk+e\bA_0(t)/\hbar)&=\sum_m\tilde\rho^{\rm eq}_{0,m}(\bk)e^{im\omega t}.
\end{align}
Together, this closed set of linear equations can in principle be solved to obtain all eigenvalues $\lambda$, and hence infer the stability of the self-consistent solutions. 

\label{sec:dm_stability_equation}

In practice, $\tilde{\boldsymbol{u}}_{\lambda}$ is high-dimensional, because it contains the density matrix at every momentum point in the Brillouin zone, and direct diagonalization of Eq.~\eqref{eq:stab_floquet_reduction} is therefore impractical. As in the End Matter, we instead eliminate $\delta \tilde{\boldsymbol P}_{\lambda}$ and $\delta \tilde{\boldsymbol{\rho}}_{\lambda}$ in favor of $\delta \tilde{\boldsymbol A}_{\lambda}$, which yields a nonlinear characteristic equation for $\lambda$ in terms of the exact Schur complement of $\addFN{  \tilde{\mathcal M}}_{\rm F}$,
\begin{equation}
  \tilde{\mathcal{G}}(\lambda) \delta \tilde{\boldsymbol A}_{\lambda} = \frac{1}{\kappa} (\lambda I+i\Omega)^2 \delta \tilde{\boldsymbol A}_{\lambda} ,
 \label{eq:stab_schur_complement}
\end{equation}
where
\begin{equation}
    \tilde{\mathcal{G}}(\lambda)=\tilde{\mathcal C}
  [\lambda I-(-i\Omega+\tilde{\mathcal K})]^{-1}\tilde{\mathcal B}.
\end{equation}
By identifying the right hand side of Eq.~\eqref{eq:stab_schur_complement} as the current [using the Maxwell relation $\boldsymbol{j}(t) = \frac{1}{\kappa} \ddot{\boldsymbol{A}}(t)$], we recognize $\tilde{\mathcal G}(\lambda)$ as the susceptibility of the current to a perturbation of the vector potential. The matrix-valued  susceptibility $\tilde{\mathcal G}(\lambda)$ can be understood as containing the Fourier components of the  Laplace transform of the  memory kernel generating the current response, $\mathcal G(t,t')$, via $\delta \bj(t) %
= \int dt'\,\mathcal G(t,t')\,\delta \boldsymbol A(t')$. In particular, by using the eigenmode solution $\delta \bA(t) = e^{\lambda t} \sum_m \delta \tilde{\bA}_{\lambda,m} e^{im\omega t}$ and $\delta \bj(t) = e^{\lambda t} \sum_n \delta \tilde{\bj}_{\lambda,n} e^{in\omega t}$, we can write $\delta \tilde{\bj}_{\lambda,n} = \sum_m [\tilde{\mathcal G}(\lambda)]_{nm} \delta \tilde{\bA}_{\lambda,m}$ where
\begin{equation}
    [\tilde{\mathcal G}(\lambda)]_{nm} = \frac{1}{T_{\mathrm{dr}}} \int_0^{T_{\mathrm{dr}}} dt \int_{-\infty}^t dt' \ e^{-(\lambda + in\omega)t } \mathcal G(t,t') e^{(\lambda + im \omega)t'}.
\end{equation}

We find the roots to Eq.~\eqref{eq:stab_schur_complement} through Newton-Raphson iteration on a box with $-0.115\leq\operatorname{Re}\lambda/\omega\leq0.05$ and $|\operatorname{Im}\lambda /\omega | \leq 3$, using the argument principle to check that we have identified all possible roots in the searched region~\cite{DelvesLyness1967}. The search is confined to finite boxes in the complex-$\lambda$ plane, and therefore does not by itself establish completeness. To count all unstable roots, we set $z=\lambda/\omega$ and construct a large half-semicircle contour, centered at the origin, that provably contains every root with $\operatorname{Re}z\geq0$. The argument principle then counts the roots inside this contour. Its radius is determined by a uniform bound on the susceptibility, which we derive in Sec.~\ref{sec:dm_susc_bound} {for the case of azimuthal symmetry. Beyond this case,  we check numerically that the number of roots do not increase as we expand the search region.}

\section{Limits on the root location}
{Here we prove that our root search from Sec.~\ref{sec:stability} yields all the roots of Eq.~\eqref{eq:stab_schur_complement}, in the case where the driven WSM problem has azimuthal symmetry.} 
We do this by establishing a bound on the susceptibility  matrix that restricts the possible root locations.

{We recall that we may restrict} the characteristic equation in Eq.~\eqref{eq:stab_schur_complement} to the two channels with harmonics $n=-1$ and $n=+1$ for Weyl nodes with azimuthal symmetry about the laser propagation axis. Writing $z=\lambda/\omega$, %
the root equation from Eq.~\eqref{eq:stab_schur_complement} can then be written as   %
\begin{equation} \label{eq:rooteq}
    \mathsf (\lambda I +i  \Omega)^2[I+{\tilde{\mathcal G}}'(\lambda/\omega)]\delta \tilde {\boldsymbol A}_{\lambda} = 0. %
\qquad
{\tilde{\mathcal G}}'(z)
=-\kappa(z\omega I +i  \Omega)^{-2}\tilde{\mathcal G}(z\omega ).
\end{equation}
The solutions for $\lambda$ must thus be located at points  where the determinant of the matrix $ \mathsf (\lambda I +i  \Omega)^2[I+{\tilde{\mathcal G}}'(\lambda)]$ vanishes. %
Since we have restricted to the fundamental harmonics, the determinant of this matrix becomes
\[
D(z)=(z^2+1)^2\det[I+{\tilde{\mathcal G}}'(z)].
\]
Where $z=\lambda/\omega$ and we have factored out $\omega$. If $z$ is an unstable physical root (with $\Re(z)>0$), the second factor must vanish, and we therefore must have %
that
\begin{equation}
\|{\tilde{\mathcal G}}'(z)\|_2\geq 1,
\end{equation}
where we have defined the operator norm $\|M\|_2\equiv\sup_{\|a\|_2=1}\|Ma\|_2$ for a matrix $M$, where $a$ is a vector and $\|a\|_2 = \sqrt{a^\dagger a}$ is the corresponding norm of a vector. Any point where $\|{\tilde{\mathcal G}}'(z)\|_2<1$ therefore cannot be a physical unstable root.  

We  now assume a fixed  upper bound on 
$\|\tilde{\mathcal G}(z)\|_2$, denoted $\overline G_0$, %
that holds throughout the right half-plane. This allows us to exclude a $z$ region from holding a physical unstable root, since
$\|(\lambda I +i  \Omega)^{-2}\|_2=1/[\omega^2d(z)^2]$, where
$d(z)=\min\{|z-i|,|z+i|\}$, and thus
\begin{equation} \label{eq:radiusrg}
\|{\tilde{\mathcal G}}'(z)\|_2
\leq\frac{\kappa\overline G_0}{\omega^2d(z)^2}
=\left[\frac{r_G}{d(z)}\right]^2,
\qquad
r_G=\frac{\sqrt{\kappa\overline G_0}}{\omega}.
\end{equation}
A root in the right half-plane must therefore satisfy $d(z)\leq r_G$, i.e., lie within a distance $r_G$ of $z=+i$ or of $z=-i$.

Our approach is to perform an exhaustive root search within a contour that fully encloses the region $d(z)\leq r_G$. We use a half-disk centered at $z=0$, %
whose boundary consists of the vertical diameter $z=iy$, with $|y|\leq L$, and the right-hand semicircle $z=Le^{i\vartheta}$, with $-\pi/2\leq\vartheta\leq\pi/2$. The radius $L$ is chosen such that every point on or outside the semicircle lies farther than $r_G$ from both $z=+i$ and $z=-i$. Since $z=\pm i$ are both a unit distance from the origin, a convenient choice for the radius of the semicircle is
\begin{equation} \label{eq:radiusL}
L=1+ r_G.
\end{equation}

\subsubsection{Derivation of the susceptibility bound} \label{sec:dm_susc_bound}

We now derive the bound $\overline G_0$ for the susceptibility matrix. Our strategy is to establish two bounds: a bound on the current response $\delta\tilde{\bj}_\lambda$ to a density-matrix perturbation $\delta\rho_\lambda(\bk,t)$,
\begin{equation}
    \|\delta\tilde{\bj}_\lambda\|_2=\|\tilde{\mathcal G}(\lambda)\,\delta\tilde{\boldsymbol A}_\lambda\|_2
    \leq C_1\int\frac{d^3\boldsymbol k}{(2\pi)^3}\,\|\delta\rho_\lambda(\bk,t)\|_2,
    \label{eq:stab_current_bound}
\end{equation}
with a constant $C_1$, and a bound on the density-matrix response to a vector potential perturbation $\delta\tilde{\boldsymbol A}_\lambda$,
\begin{equation}
    \|\delta\rho_\lambda(\bk,t)\|_2
    \leq b(\bk)\,\|\delta\tilde{\boldsymbol A}_\lambda\|_2
    \qquad{\rm for}\qquad \operatorname{Re}z\geq0,
    \label{eq:stab_density_bound}
\end{equation}
with a $z$-independent function $b(\bk)$. Here, $\delta\tilde{\bj}_\lambda=(\ldots,\delta\tilde{\bj}_{\lambda,-1},\delta\tilde{\bj}_{\lambda,0},\delta\tilde{\bj}_{\lambda,1},\ldots)^T$ collects the Fourier coefficients of the current response. Above, we have generalized the norm $\|X(t)\|_2^2=\langle X(t),X(t)\rangle$ for time-periodic quantities $X(t)$, using the inner product $\langle X,Y\rangle=T_{\rm dr}^{-1}\int_0^{T_{\rm dr}}dt\,\mathrm{Tr}[X^\dagger(t)Y(t)]$.  We choose this definition so that, by Parseval's theorem, $\| X(t) \|_2 = \|\tilde{X} \|_2$, where $\tilde{X}$ denotes the vector of the Fourier harmonics of $X(t)$.

Combining the two bounds and using $\|\tilde{\mathcal G}(z)\|_2=\sup_{\|\delta\tilde{\boldsymbol A}_\lambda\|_2=1}\|\tilde{\mathcal G}(z)\,\delta\tilde{\boldsymbol A}_\lambda\|_2$ gives
\begin{equation}
\left\|\tilde{\mathcal G}(z)\right\|_2
\leq
C_1\int
\frac{d^3\boldsymbol{k}}{(2\pi)^3}
\,b(\bk)
\equiv
\overline{G}_{0}
\qquad{\rm for}\qquad \operatorname{Re}z\geq0.
 \label{eq:stab_G_bound}
\end{equation}
It remains to determine $C_1$ and $b(\bk)$. Below, we first establish Eq.~\eqref{eq:stab_current_bound} with $C_1=\sqrt2\,ev_F$, and then determine $b(\bk)$ using the linearized master equation.

To compute $C_1$, we first note that because $\mathcal C(t)$ is linear and periodic, the current response can be written as $\delta\bj(t)=e^{\lambda t}\delta\bj_\lambda(t)$, where
\begin{equation*}
    \delta\bj_\lambda(t)
    =-e\int\frac{d^3\boldsymbol k}{(2\pi)^3}\,\mathrm{Tr}\!\left[\delta\rho_\lambda(\bk,t)\,\boldsymbol v_\perp(\bk,t)\right],
\end{equation*}
and $\boldsymbol v_\perp=(v_x,v_y)$, with $v_a=\hbar^{-1}\partial_{k_a}h(\bk,t)$ for $a=x,y$ denoting the in-plane velocity operator. The bound follows from 
\begin{equation*}
    \|\delta\tilde{\bj}_\lambda\|_2=\|\delta\bj_\lambda(t)\|_2
    \;\leq\;e\int\frac{d^3\boldsymbol k}{(2\pi)^3}\,
    \big\|\mathrm{Tr}[\delta\rho_\lambda(\bk,t)\,\boldsymbol v_\perp(\bk,t)]\big\|_2
    \;\leq\;\sqrt2\,ev_F\int\frac{d^3\boldsymbol k}{(2\pi)^3}\,\|\delta\rho_\lambda(\bk,t)\|_2.
\end{equation*}
Here we have used Bessel's inequality: since $\mathrm{Tr}[v_a^\dagger v_b]=2v_F^2\delta_{ab}$, $\|\mathrm{Tr}[\delta\rho_\lambda(\bk,t)\,\boldsymbol v_\perp(\bk,t)]\|_2\leq\sqrt2\,v_F\|\delta\rho_\lambda(\bk,t)\|_2$. This establishes the inequality in Eq.~\eqref{eq:stab_current_bound} with $C_1=\sqrt2\,ev_F$.

We next turn to the function $b(\bk)$ that bounds the density-matrix response, Eq.~\eqref{eq:stab_density_bound}, and begin by using azimuthal symmetry to simplify our analysis. Azimuthal symmetry restricts the vector potential perturbation to the fundamental harmonic, allowing it to be written as [see Eq.~\eqref{eq:rooteq}],
\begin{equation}
    \delta\boldsymbol A_\lambda(t)=a_{\lambda,+}\,\boldsymbol u_+(t)+a_{\lambda,-}\,\boldsymbol u_-(t),
    \qquad
    \boldsymbol u_+(t)=\frac{\hat{\boldsymbol x}+ic_\eta\hat{\boldsymbol y}}{2}\,e^{-i\omega t},
    \quad
    \boldsymbol u_-(t)=\boldsymbol u_+(t)^*.
    \label{eq:stab_circular_channels}
\end{equation}
To evaluate $\delta\rho_\lambda(\bk,t)$, we insert $\delta\rho(\bk,t)=e^{\lambda t}\delta\rho_\lambda(\bk,t)$ and $\delta\boldsymbol A(t)=e^{\lambda t}\delta\boldsymbol A_\lambda(t)$ into the linearized master equation, $\partial_t\delta\rho=\mathcal B(t)[\delta\boldsymbol A]+\mathcal K(t)[\delta\rho]$, and use the definition for $\mathcal K(t)$ in Eq.~\eqref{eq:k(t)} to obtain
\begin{equation}
    (\lambda+\partial_t)\,\delta\rho_\lambda(\bk,t)
    =\frac{i}{\hbar}\left[\delta\rho_\lambda(\bk,t),h(\bk,t)\right]
    -\frac{\delta\rho_\lambda(\bk,t)}{\tau}
    +\mathcal B(t)\big[a_{\lambda,+}\boldsymbol u_++a_{\lambda,-}\boldsymbol u_-\big](\bk).
    \label{eq:stab_periodic_lindblad}
\end{equation}
To bound $\|\delta\rho_\lambda(\bk,t)\|$, we take the inner product of both sides of Eq.~\eqref{eq:stab_periodic_lindblad} with $\delta\rho_\lambda$ itself, which yields the identity
\begin{align}
    \lambda\,\|\delta\rho_\lambda(\bk,t)\|^2
    +\big\langle\delta\rho_\lambda(\bk,t),\partial_t\delta\rho_\lambda(\bk,t)\big\rangle
    =&\;\frac{i}{\hbar}\big\langle\delta\rho_\lambda(\bk,t),[\delta\rho_\lambda(\bk,t),h(\bk,t)]\big\rangle
    -\frac{\|\delta\rho_\lambda(\bk,t)\|^2}{\tau}
    \notag\\
    &+\big\langle\delta\rho_\lambda(\bk,t),\,\mathcal B(t)[a_{\lambda,+}\boldsymbol u_++a_{\lambda,-}\boldsymbol u_-](\bk)\big\rangle.
    \label{eq:stab_scalar_identity}
\end{align}
We now simplify Eq.~\eqref{eq:stab_scalar_identity} term by term. Since the quantity we seek to bound, $\|\delta\rho_\lambda(\bk,t)\|^2$, is real, no information is lost by taking the real part of both sides---and doing so simplifies the identity considerably. Several terms vanish. The second term on the left, $\langle\delta\rho_\lambda(\bk,t),\partial_t\delta\rho_\lambda(\bk,t)\rangle$, is zero: its real part is the time average of the derivative $\frac12\partial_t\mathrm{Tr}[\delta\rho_\lambda^\dagger(\bk,t)\,\delta\rho_\lambda(\bk,t)]$, which vanishes for a periodic function. The commutator term, $\frac{i}{\hbar}\langle\delta\rho_\lambda(\bk,t),[\delta\rho_\lambda(\bk,t),h(\bk,t)]\rangle$, is purely imaginary: for any matrix $X$ and Hermitian $h$, $\mathrm{Tr}\{X^\dagger[X,h]\}=\mathrm{Tr}\{(X^\dagger X-XX^\dagger)h\}$ is real, so the prefactor $i/\hbar$ renders the term imaginary. Finally, the real part of the last term, $\langle\delta\rho_\lambda(\bk,t),\mathcal B(t)[a_{\lambda,+}\boldsymbol u_++a_{\lambda,-}\boldsymbol u_-](\bk)\rangle$, can be bounded using the Cauchy--Schwarz inequality,
\begin{equation}
    \operatorname{Re} \langle\delta\rho_\lambda(\bk,t),\mathcal B(t)[a_{\lambda,+}\boldsymbol u_++a_{\lambda,-}\boldsymbol u_-](\bk)\rangle \leq\|\delta\rho_\lambda(\bk,t)\|\,
    \big\|\mathcal B(t)[a_{\lambda,+}\boldsymbol u_++a_{\lambda,-}\boldsymbol u_-](\bk)\big\|.
\end{equation}
Rearranging, we obtain
\begin{equation}
    \|\delta\rho_\lambda(\bk,t)\|^2
    \leq \left(\operatorname{Re}\lambda+\frac{1}{\tau}\right)^{-1} \|\delta\rho_\lambda(\bk,t)\|\,
    \big\|\mathcal B(t)[a_{\lambda,+}\boldsymbol u_++a_{\lambda,-}\boldsymbol u_-](\bk)\big\|.
    \label{eq:stab_dissipativity}
\end{equation}
Since $\mathcal B(t)$ is linear, 
\begin{equation*}
    \big\|\mathcal B(t)[a_{\lambda,+}\boldsymbol u_++a_{\lambda,-}\boldsymbol u_-](\bk)\big\|
    =\big\| a_{\lambda,+} \mathcal B(t)[\boldsymbol u_+](\bk)+a_{\lambda,-}\mathcal B(t)[\boldsymbol u_-](\bk)\big\|.
\end{equation*}
Applying the triangle inequality, 
\begin{equation*}
    \big\|\mathcal B(t)[a_{\lambda,+}\boldsymbol u_++a_{\lambda,-}\boldsymbol u_-](\bk)\big\|
    \leq\Big[\big\|\mathcal B(t)[\boldsymbol u_+](\bk)\big\|^2+\big\|\mathcal B(t)[\boldsymbol u_-](\bk)\big\|^2\Big]^{1/2}
    \big(|a_{\lambda,+}|^2+|a_{\lambda,-}|^2\big)^{1/2}.
\end{equation*}
Finally, for $\operatorname{Re}\lambda\geq0$ (i.e., $\operatorname{Re}z\geq0$), we have $\operatorname{Re} \lambda + 1/\tau \geq 1/\tau$. Furthermore, we can replace $(|a_{\lambda,+}|^2+|a_{\lambda,-}|^2)^{1/2}=\sqrt2\,\|\delta\tilde{\boldsymbol A}_\lambda\|_2$. Substituting these results into Eq.~\eqref{eq:stab_dissipativity}, and dividing both sides by $\|\delta\rho_\lambda(\bk,t)\|$, we obtain the inequality Eq.~\eqref{eq:stab_density_bound} with
\begin{equation}
    b(\bk)=\sqrt2\,\tau\Big[\big\|\mathcal B(t)[\boldsymbol u_+](\bk)\big\|^2+\big\|\mathcal B(t)[\boldsymbol u_-](\bk)\big\|^2\Big]^{1/2},
    \label{eq:stab_bk}
\end{equation}
which is entirely $z$-independent.

With $C_1=\sqrt2\,ev_F$ and $b(\bk)$ given by Eq.~\eqref{eq:stab_bk}, Eq.~\eqref{eq:stab_G_bound} yields the uniform bound
\begin{equation*}
    \overline G_0=2ev_F\int\frac{d^3\boldsymbol k}{(2\pi)^3}\,\tau
    \Big[\big\|\mathcal B(t)[\boldsymbol u_+](\bk)\big\|^2+\big\|\mathcal B(t)[\boldsymbol u_-](\bk)\big\|^2\Big]^{1/2}
\end{equation*}
for the right-half plane $\operatorname{Re} \lambda \geq 0$. In our calculations, this integral is evaluated numerically. This bound allows us to compute the radius $L$ of the semicircle enclosing all possible unstable roots [see Eqs.~\eqref{eq:radiusL} and~\eqref{eq:radiusrg}]. We then use the argument principle to count all roots in this semicircle contour, and verify that it matches the number of unstable roots found via Newton-Raphson iteration.

\bibliographystyle{apsrev4-1}
\bibliography{references_use}